\documentclass[preprint]{aastex}
\usepackage{ifpdf}
\usepackage{graphicx}
\usepackage{natbib}

 \def\gsim{ \lower .75ex \hbox{$\sim$} \llap{\raise .27ex \hbox{$>$}} }
\def\lsim{ \lower .75ex \hbox{$\sim$} \llap{\raise .27ex \hbox{$<$}} }

\begin{document}			% REQUIRED

\newcommand{\cSIStwo}{cSIS$_{2{\rm kpc}}$}
\newcommand{\cSISfour}{cSIS$_{4{\rm kpc}}$}
\newcommand{\NFWfour}{NFW$_{4{\rm kpc}}$}

\title{Tests of Modified Gravity with Dwarf Galaxies}

\author{Bhuvnesh Jain\altaffilmark{1} and Jake VanderPlas\altaffilmark{2}}
\bibliographystyle{apj}

\altaffiltext{1}{Department of Physics \& Astronomy, University of Pennsylvania, 
Philadelphia, PA 19104} 
\altaffiltext{2}{Astronomy Department, University of Washington, Seattle, WA 98195}

%\date{\today}		 		% your own text, a date, or \today
% --------------------- end of the preamble ---------------------------
%\bibliographystyle{apj}

\begin{abstract}
In modified gravity theories that seek to explain cosmic acceleration, 
dwarf galaxies in low density environments 
can be subject to enhanced forces. The class of scalar-tensor 
theories, which includes $f(R)$ gravity, predict such a force enhancement 
(massive galaxies like the Milky Way can evade it through a screening mechanism 
that protects the interior of the galaxy  from this ``fifth'' 
force). We study observable deviations from GR in the disks of late-type 
dwarf galaxies moving under gravity. 
The fifth-force acts on the dark matter and HI gas disk, but
not on the stellar disk owing to the self-screening of main sequence stars. 
We find four distinct observable effects in such disk galaxies: 
1. A displacement of the stellar disk from the HI disk. 
2. Warping of the stellar disk along the direction of the external force. 
3. Enhancement of the rotation curve measured from the HI 
gas compared to that of the stellar disk. 
4. Asymmetry in the rotation curve of the stellar disk. 
We estimate that the spatial effects can be up to 1 kpc and the rotation
velocity effects about 10 km/s in infalling dwarf galaxies. 
Such deviations are measurable: we expect that
with a careful analysis of a sample of nearby 
dwarf galaxies one can improve astrophysical constraints 
on gravity theories by over three orders of magnitude, and even solar 
system constraints by one order of magnitude. Thus effective tests of 
gravity along the lines suggested by 
\citet{Hui09} and \citet{Jain11} can be carried out  with low-redshift galaxies,  
though care must be exercised in understanding
possible complications from astrophysical effects.  
\end{abstract}

\section{Introduction}
\label{sec:introduction}
The true nature of the observed accelerated expansion of the universe 
remains a mystery, but the possible
solutions to this puzzle fall into two broad categories.  The first 
is to posit the existence of a new component
of the universe with an appropriate equation of state to cause the observed
acceleration.  This ``dark energy'' may be due to an exotic particle or field,
or be related to the vacuum energy of space itself.  A second approach is to
seek to explain this acceleration through modifications to the field 
equations of general relativity (GR) itself. In recent years this modified
gravity (MG) approach has received attention and different approaches
are being actively developed -- for a review see e.g. \citet{Jain10}.

%GR is an extremely successful theory, tested to high precision over a large
%range of length scales.  Hence any tenable modification must be suppressed
%within the solar system.  [Talk briefly about chameleon, DGP, Vainstein, etc]
%Until recently, observable predictions of modified gravity (MG) 
%were cosmological in nature, e.g. in the clustering 
%of galaxies at scales $>10$Mpc.  Recently,
%however, it has been recognized that dwarf galaxies in cosmic voids may
%provide a unique testing ground to put useful constraints on MG parameters
%(Hui, Nicolis \& Stubbs 2009, Chang \& Hui 2011, Davis et al. 2011).  
%Here we build on this work, and argue
%that observations of rotation curves and morphologies of a handful of
%suitable dwarf galaxies could constrain modified gravity to a precision 
%competetive with that of an LSST-scale survey of structure.

%Weinberg (1965) argued that general requirements based on Lorentz invariance lead to 
%GR being the unique large-scale description of a gravitational force mediated by a 
%massless spin-2 field. Hence scalar-tensor theories, in which an additional scalar field 
%couples to the Ricci scalar in the action, are expected to be the generic means of modifying
%gravity on cosmological scales (exceptions include the DGP model 
%in which the graviton is a resonance state, or massive gravity theories). 

%\subsection{Modified Gravity through a Screened Fifth Force}

A modification of GR on large (astrophysical) scales generically 
leads to scalar-tensor theories of gravity, where a new scalar 
field couples to gravity. Equivalently
these theories can be described via a coupling of the scalar field  
to matter, which  leads to enhancements of the gravitational force. 
Nonrelativistic matter -- such as the stars, gas, and dust in galaxies --
will feel this enhanced force, which in general lead to larger dynamically
inferred masses.  The discrepancy can be up to a factor of 1/3 in $f(R)$ 
or DGP gravity. 
%(see Jain \& Khoury 2010 for a review). 
We note that photons,
being relativistic, do not feel the enhanced force, so that lensing 
probes the true mass distribution.

%Isolated nonrelativistic objects such as stars and galaxies experience the 
%enhanced forces leading to higher gravitationally induced velocities.  
%While lensing masses are true masses (light deflection is not altered), 
%dynamical masses, i.e. the masses inferred from the motion of stars or 
%galaxies around the object, are larger. The discrepancy can be up to a 
%factor of 1/3 in $f(R)$ or DGP gravity (see Jain \& Khoury 2010 for a review). 

This enhanced gravitational force should be detectable through  
fifth force  experiments, 
tests of the equivalence principle (if the scalar coupling to matter 
varied with the properties of matter) or through the 
orbits of planets around the Sun \citep{Will05}. %(Will 2005). 
\citet{Khoury04} %Khoury \& Weltman (2004) 
proposed that nonlinear screening of the 
scalar field, called chameleon screening, can suppress the fifth force 
in high density environments such as the Milky Way, so that Solar System and 
lab tests can then be satisfied. This screening was originally suggested 
to hide the effects of a quintessence-like scalar that forms the dark energy 
and may couple to matter (generically such a coupling is expected unless 
forbidden by a symmetry). Hence there are reasons to expect such a screening 
effect to operate in either a dark energy or modified gravity scenario. Indeed, since
there are only a handful of screening mechanisms (Vainshtein and symmetron 
screening are discussed below), small scale tests of gravity 
that rely on distinct signatures of screening are useful discriminators of 
cosmological models. 

Dwarf galaxies in low-density environments may remain unscreened as the 
Newtonian potential $\Phi_N$, which determines the level of screening, 
is an order of magnitude smaller in magnitude than in the Milky Way. Hence 
dwarf galaxies can exhibit manifestations of modified forces in both 
their infall motions and internal dynamics. 
\citet{Hui09} %Hui, Nicolis and Stubbs (2009) 
discuss various observational effects, in particular the fact that 
rotation velocities of HI gas can be enhanced. Interestingly, 
stars can self-screen so that the stellar disk in an unscreened 
dwarf galaxy may have lower rotational velocity than the HI disk. 
Indeed  the small scale dynamics 
of nearby galaxies may have a bigger signal of MG than 
large-scale perturbations -- see \citet{Jain11}
for a discussion of observational approaches. 

In this work, we discuss observable effects on disk galaxies that can arise due 
to their interaction with a neighbor or other sources of an external 
gravitational field. We focus on late-type dwarf galaxies as these 
are most likely to be unscreened. The basic effect is that if the 
stellar disk remains screened, it will lag the dark matter and HI 
disk in the infall towards another galaxy. This may lead to a 
separation of the stellar disk from the center of mass 
of the dark matter and from the HI disk. Distortions of the morphology 
and rotation curves of the stellar disk may result. Our goal is
to estimate the size of these effects. 
A similar effect in a somewhat different context was discussed by
\citet{Kesden06}, %Kesden \& Kamionkowski (2006), 
who used tidal effects on dwarf galaxies to 
test the equivalence principle in the Milky Way. They were interested in 
tests that may reveal an additional force that acts only between dark matter 
particles, as were  
\citet{Gradwohl92}. %Gradwohl \& Frieman (1992). 
We are 
concerned with a universal force but one that may be suppressed depending 
on the density of the tracer. We discuss the connections and differences of
 our work from these papers below. 

While we do not wish to be restricted to particular models, we do 
work within a modified gravity
scenario that relies on chameleon screening 
(we discuss below how the recently proposed symmetron and, in some 
situations, Vainshtein screening mechanisms 
%\citep{Hinterbichler10} %(Hinterbichler and Khoury 2010)
 are also testable via the effects we discuss). Further we assume that sufficiently small 
galaxies, of virial masses $M\sim 10^{10}-10^{11} M_\odot$ are unscreened. 
The virial radius and peak circular velocity (or velocity dispersion) 
scale with halo mass as $r_{\rm vir}, v_c \propto M^{1/3}$. 
So the Newtonian potential at 
the virial radius scales roughly as 
$|\Phi_N| \propto M/r_{\rm vir} \propto M^{2/3} \propto v_c^2$. 
The Milky Way has $|\Phi_N| \approx 10^{-6}$ (with $c=1$), so a dwarf galaxy 
with a rotation speed of about $50$ km/s (a factor of 4 smaller 
than the Milky Way) will have $|\Phi_N| \lsim 10^{-7}$. 
We will assume that such dwarf galaxies are unscreened provided 
they occupy a sufficiently low density environment. Within a galaxy 
group or cluster, the same sized dwarf galaxy would
be screened, which provides a potentially useful ``control'' sample. 
Main sequence stars happen to have $|\Phi_N| \approx 10^{-6}$, so stars 
are expected to be self-screened even inside unscreened galaxies. 

We estimate analytically the displacement of the stellar disk in a dark 
matter halo in \S\ref{sec:two_body_infall}. In \S\ref{sec:dynamics} we 
describe the distortions of the stellar disk and present results from 
simple simulations. The implications and caveats associated with these 
results are discussed in \S\ref{sec:discussion}. We conclude in 
\S\ref{sec:conclusion}.

\section{Two-body infall}
\label{sec:two_body_infall}

\begin{figure}[t,h]
\centering
\includegraphics[width=8cm]{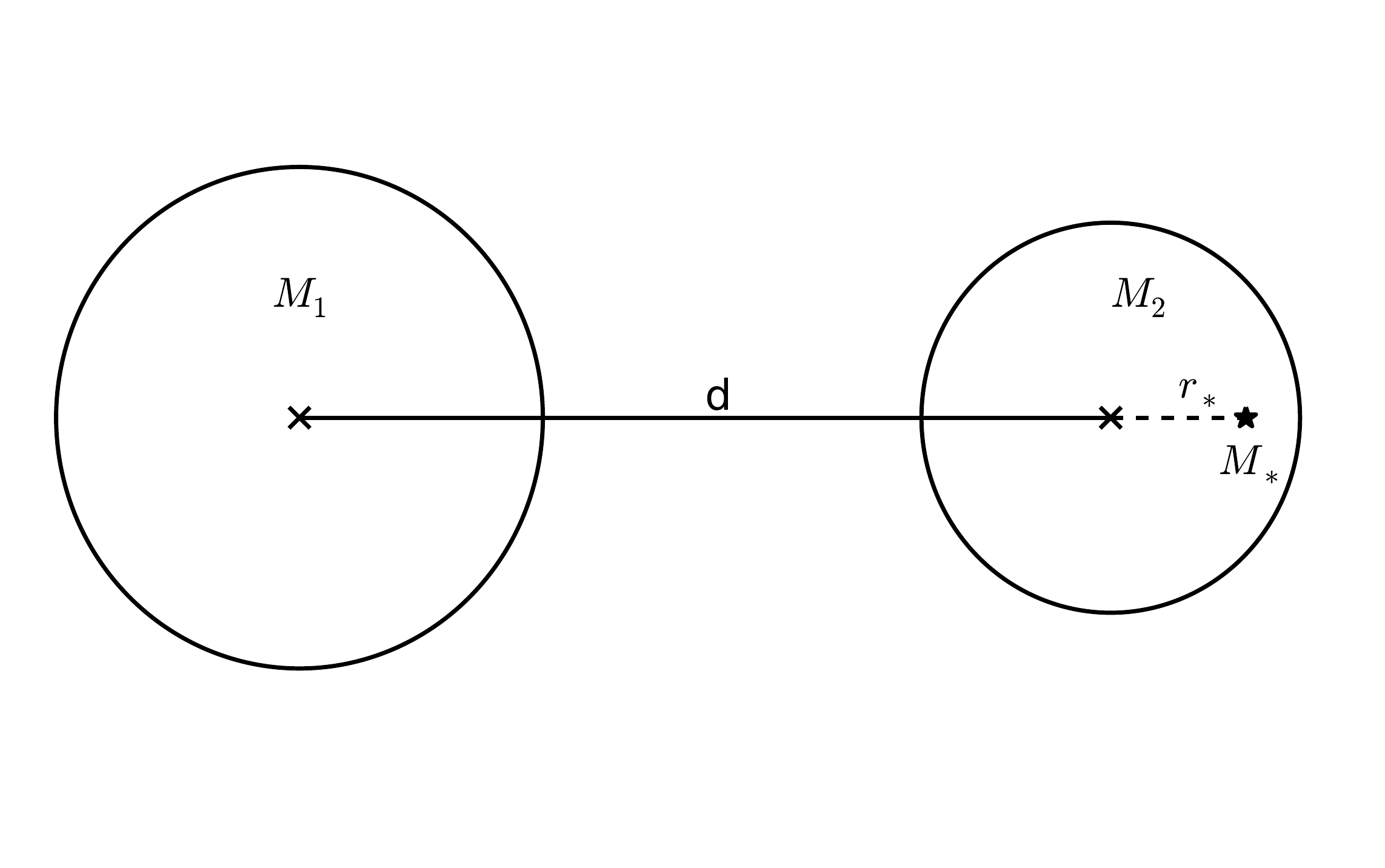}
\caption{We consider the infall of a dwarf galaxy with mass $M_2$ towards another galaxy of larger
mass $M_1$. The stars lag the center of mass of $M_2$ by distance $r_\star$. 
  \label{fig:twohalos}
}
\end{figure}

Consider a dwarf galaxy in an external gravitational field: for concreteness we will consider 
the force due to a neighboring galaxy, as shown in Figure~\ref{fig:twohalos}. We are interested in the 
effect of the attractive ``fifth" force due to a scalar field in the context of a chameleon-type screening
mechanism that leaves objects with Newtonian potential $|\Phi_N|  \lsim 10^{-7}$ unscreened. 
Since stars in dwarf galaxies are able to self-screen, the stellar disk can separate from 
the center of mass of the dark matter (DM) halo of the 
dwarf galaxy because  the fifth force acts only on the DM. 

Figure~\ref{fig:twohalos} shows dwarf galaxy of mass $M_2$ falling 
towards another galaxy of  $M_1>M_2$ with
$d$ being the instantaneous separation between their centers of mass.  
The virial radii (not labeled above) are $r_1$ and $r_2$. Within $M_2$, 
the dark matter feels an enhanced force towards $M_1$, while the stars 
feel only GR. As the halos approach each other, the stars fall behind the 
DM with a separation that increases with time. At the moment when the 
galaxies are at distance $d$, let the stellar component of mass $M_\star$ 
lag the DM by distance $r_\star$. Let's make the following simplifications: 
\begin{enumerate}
\item The DM dominates $M_2$ and the self-gravity of the stars is negligible. This is 
valid for most dwarf galaxies outside of the innermost $\lsim 1$ kpc region. 
%\item The stellar disk or bulge is smaller than $r_\star$, i.e. we may regard the stars as a point 
%mass at distance $r_\star$ from the center of mass of $M_2$ for purposes of estimating $r_\star$. 
\item The force enhancement for the DM is well approximated by changing the gravitational constant to $G>G_N$. This is valid provided the dwarf galaxy is in an unscreened environment so that the scalar field is nearly massless. 
\item We will also assume $r_\star/d \ll 1$. While we are interested in cases where this ratio is not
negligibly small,  order unity deviations are easily excluded observationally. 
\end{enumerate}

\subsection{Displacement of the stellar component}
\label{sec:displacement}

Our starting point is to seek the minimum of the gravitational potential
felt by the stellar disk in the reference frame defined by halo 2.  
The disk will rotate about this minimum, which makes it a useful estimate
of the overall displacement $r_\star$.  We will consider a star near the center
of the disk here -- in \S\ref{sec:dynamics}, 
we apply a more rigorous approach which takes into account the dynamics
of a finite-sized disk.

The acceleration of $M_2$ towards $M_1$ is given by
\begin{equation}
  a_2 = \frac{G M_1}{d^2}
\end{equation}
In this accelerating reference frame, the instantaneous
potential felt by a star at location $r_\star$ is given by
\begin{equation}
  \label{eqn:phi_def}
  \Phi(r_\star) = \Phi_2(r_*)
  - \frac{G_N M_1}{d+r_\star} - \frac{G M_1}{d^2}r_\star
\end{equation}
where $\Phi_2$ is the potential due to the dark matter halo
of the infalling galaxy, the second term is the effect of $M_1$ on the star 
and the final term accounts for 
the accelerating reference frame.  
Minimizing this with respect to $r_\star$, in the limit $r_\star/d \ll 1$, 
gives
\begin{equation}
  \label{eqn:infall_relation}
  \frac{M_1}{d^2} \left(\frac{G-G_N}{G_N} +\frac{2r_\star}{d}\right) = \frac{M_2(<r_\star)}{r_\star^2} 
\end{equation}
Where we have used ${\rm d}\Phi_2/{\rm d}r = G_NM_2(<r)/r^2$.
Care is needed here: Equation~\ref{eqn:infall_relation} describes not only the
desired local minimum of the potential, but also a possible local maximum which
is not a physical solution.  The solution is a minimum if the
second derivative of Equation~\ref{eqn:phi_def} is positive. 
%This yields the 
%following criterion which must be met by a physical solution:
%\begin{equation}
 % \label{eqn:minimum_criterion}
 % \left(\frac{r_\star}{2}\right)
 % \frac{\mathrm{d}}{\mathrm{d}r_\star}M_2(<r_\star) 
 % > M_2(<r_\star) + \left(\frac{r_\star}{d}\right)^3M_1
%\end{equation}

Let us introduce the dimensionless parameters 
\begin{equation}
  \label{eqn:params}
  m \equiv M_2/M_1 , \ 
  g \equiv \Delta G/G_N = \frac{G-G_N}{G_N}, \ 
  \mu \equiv r_2/d, \ 
  \nu\equiv r_\star/r_2, \ \  {\rm such\ that}\ \  r_\star/d=\mu\nu
\end{equation}
all of which are smaller than 1, but not 
necessarily much smaller. For given values of the other parameters, we are 
interested in how large $r_\star$ is (in \S\ref{sec:dynamics} 
we also consider the warping effect 
on a finite disk and on rotation velocities): values of order a kpc 
would be potentially observable.  We also note that typical values for the 
other scales are: $r_2\approx 20-50$ kpc and $d \gsim 100$ kpc.

Writing Equation~\ref{eqn:infall_relation} in terms of the 
dimensionless parameters defined in
Equation~\ref{eqn:params}, we obtain
\begin{equation}
  \label{eqn:infall_dimensionless}
  \mu^2\nu^2(g+2\mu\nu) = m\frac{M_2(<r_\star)}{M_2}
\end{equation}
%We should take a moment to remark on the expected value of each of these 
%parameters.  
We assume the halo masses satisfy $m<1$, and that the modification
to gravity is order $g\sim 1$.  If the halos have not yet collided, then we
must have $\mu\lesssim 1/2$, depending on the relative virial radii of the
halos.  Finally, we expect $\nu \ll 1$, or this effect would likely have been
observed in the past.

We can proceed by exploring a few possible forms of $M_2(<r_\star)$.
Let us first consider the case where  $M_2$ follows a power law profile: 
$\rho(r) \propto 1/r^\alpha$ so that $M_2(<r_\star)/M_2 = \nu^{3-\alpha}$.  
Equation~\ref{eqn:infall_dimensionless} can then be simplified to
\begin{equation}
  \label{eqn:infall_power_law}
  \mu^2(g+2\mu\nu) = m\nu^{1-\alpha}
\end{equation}
The solution for $\nu$ is then easy to see graphically: the LHS of 
Equation~\ref{eqn:infall_power_law} is a straight line with a positive 
slope and the 
solution is its  intersection with the RHS which 
is a power law that's either a positive or negative power.
 The negative power gives a solution for $\nu\sim 1$ that corresponds
 to a local maximum of the potential, whereas we are interested in the minimum. 
 
 The problem with a steep inner profile is that 
with $\alpha>1$, the potential diverges at $r\to 0$, so that the
approach in Equation~\ref{eqn:infall_relation} of setting the gradient to zero
will not find the true minimum.  Thus we expect that for halos with
steep inner profiles, the magnitude $r_\star$ of the mean separation 
will be negligible. The inner profile of Navarro, Frenk \& White (NFW) halos is $\alpha=1$, thus we 
expect $r_\star$ to be very small for such halos (we will see in 
\S\ref{sec:dynamics} that finite sized disks in such halos 
nevertheless show observable effects). 
%as the criterion in Eqn.~\ref{eqn:minimum_criterion} gives
%$\alpha<1 - 2(\mu^3\nu^\alpha/m)$ for the power law profile.

%Let's try two examples. 
%\begin{enumerate}
%\item
%Let $\alpha=1$, the inner slope of NFW. Then $\nu$ drops out of the 
%RHS and we can write
%\begin{equation} 
%  \nu = \frac{1}{2\mu}\left[\frac{m}{\mu^2}-g\right]
%\end{equation}
 % If we assume $g\sim 1$, $m\sim 1/8$ and $\mu\sim 1/10$,  then 
%$\nu \sim 1$ for the equation to hold.  
%This invalidates our assumption that
%$r_\star \ll d$, so this is not a physical solution.  This could be
%anticipated, as $\alpha=1$ does not meet the criterion in 
%Eqn.~\ref{eqn:minimum_criterion}.

%The problem with a steep profile is that for $\rho \propto 1/r^\alpha$ 
%with $\alpha>1$, the potential diverges at $r\to 0$, so that the
%approach in Equation 2 of setting the gradient to zero
%will not find the true minimum.  Thus we expect that for halos with
%steep inner profiles, the magnitude $r_\star$ of the mean separation 
%will be negligible.

%\item
Realistic dwarf galaxy halos can have shallower central profiles, the limit of which is
simply a flat core.  This regime can be approximated by a
constant density sphere, $\alpha=0$.  In this case we have
\begin{equation}
  \nu = \frac{\mu^2}{m}(g-2\mu\nu) \approx \frac{\mu^2 g}{m}
\end{equation}
If $g\sim 1$, $\mu\sim 1/10$, and $m\sim 1/8$, we get $\nu \sim 1/10$.
This is potentially an interesting number, as it implies $r_\star \sim 1$kpc
for typical sizes of dwarf galaxies.  Real halos have core radii on the
order of a few kpc, so this constant density approximation quickly breaks down.
To obtain a more rigorous estimate of the expected deviation 
$r_\star$ we consider more realistic density profiles. 
%\end{enumerate}

\subsection{Realistic density profiles}
The power-law profiles discussed above are overly simplistic, but they 
point to the existence of two regimes with very different qualitative 
behavior: halos with steep cores and halos with flat cores.  
To represent realistic dwarf galaxy
halos with these properties, we will use the following two halo profiles:\\
{\bf Cored Isothermal Sphere (cSIS):}
A cored isothermal sphere has a flat core which, for low mass galaxies,
extends up to $\sim 1-4$kpc from the center.  
It is a good approximation to DM halos in observed dwarf galaxies
\citep[see][]{Swaters11}. %(see Swatters et al. 2011).  
The density profile is given by
\begin{equation}
  \label{eqn:cSIS_def}
  \rho(r) = \frac{\rho_0}{1+(r/r_0)^2}
\end{equation}
Letting $x\equiv r/r_0$, the mass within a radius $r$ can be expressed
\begin{equation}
  M(<r)   = 4\pi\rho_0 r_0^3 \left[x-\tan^{-1}(x)\right]
\end{equation}
and the potential $\Phi$ felt by a star at radius $r$  is
\begin{equation}
  \label{eqn:cSIS_Phi}
  \Phi(r) = 4\pi G_N \rho_0r_0^2
  \left[\frac{\log(x^2+1)}{2} + \frac{\tan^{-1}(x)}{x}\right]
\end{equation}
\\
{\bf NFW halo:}
The NFW halo is a steeper-cored profile which is well-motivated both
observationally and theoretically.  For a suitable choice of parameters, the 
NFW and cSIS profile yield rotation curves that are reasonable fits to
 observations over relevant radii. 
The NFW density profile is given by
\begin{equation}
  \label{eqn:NFW_def}
  \rho(r) = \frac{\rho_0}{(r/r_0)(1+r/r_0)^2}
\end{equation}
Again letting $x\equiv r/r_0$, the mass within a radius $r$ can be expressed
\begin{equation}
  M(<r) = 4\pi\rho_0 r_0^3\left[\log(x+1)-\frac{x}{x+1}\right]
\end{equation}
and the potential $\Phi$ at radius $r$ is
\begin{equation}
  \label{eqn:NFW_Phi}
  \Phi(r) = 4\pi G_N \rho_0r_0^2\left[1-\frac{\log(x+1)}{x}\right]
\end{equation}

Using these two halo profile shapes in Equation~\ref{eqn:infall_dimensionless}
gives
\begin{eqnarray}
  \label{eqn:cSIS_scaling}
  \mathrm{cSIS:} &\mu^2\nu^2(g+2\mu\nu) = 
  m\frac{\nu c - \arctan(\nu c)}{c - \arctan(c)}\\
  \label{eqn:NFW_scaling}
  \mathrm{NFW:} &\mu^2\nu^2(g+2\mu\nu) = m\frac{\log(\nu c+1) - \nu c /(\nu c+1)}{\log(c+1) - c /(c+1)}
\end{eqnarray}
where we have defined the concentration parameter $c\equiv r_2/r_0$, 
where $r_2$ is the virial radius of the halo. 
%The nice thing here is that
%the particular choice of halo parameters, i.e. $r_0$ and $\rho_0$, cancel
%out so that 
%%BJ: since the solution is in terms of r*/r2, it does depend on the halo profile. 
%%so not really indepednent of r0 and rho0. 

\begin{figure}[t]
\centering
\includegraphics[width=9.5cm]{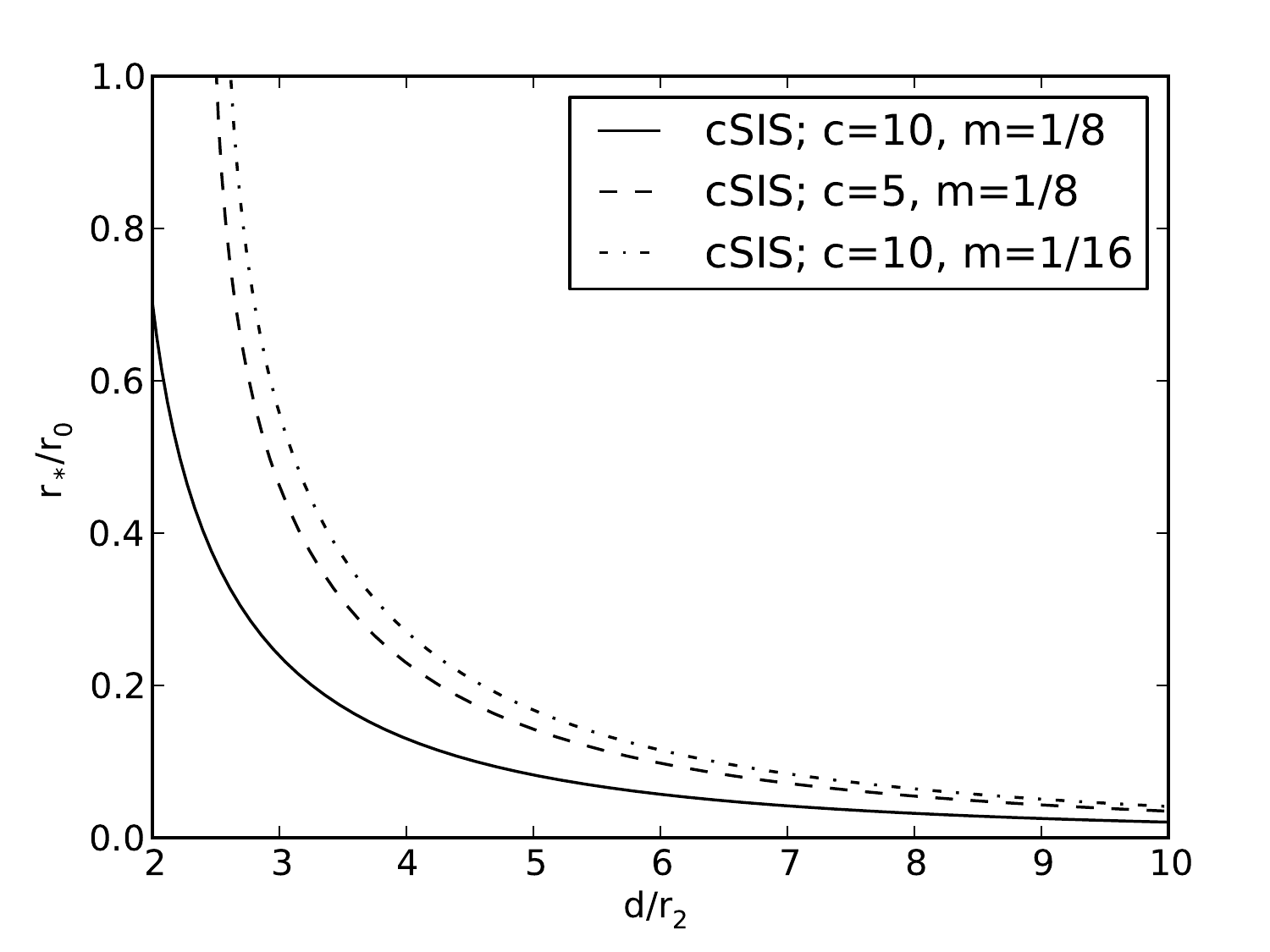}
\caption{
The displacement of stars $r_\star/r_0$ as a function of halo separation
for the cSIS halo profile. The curves are solutions of 
Equation~\ref{eqn:cSIS_scaling}.
%The solid curve uses the fiducial values $c=10$ and $m=1/8$, the dotted curve 
%has  $c=20$ and the dashed curve $m=1/16$.  
With $r_0\gsim 1$kpc, values of $r_\star/r_0 > 0.1$ are potentially 
observable for nearby dwarf galaxies. 
}
\label{fig:profile}
\end{figure}

We can determine the expected size of $r_\star$ for choices 
of the other parameters in the equation.  
As discussed above, the NFW profile does not give a true solution  
for finite $r_\star$ -- we will see below that observable effects with such 
a cuspy profile arise only 
when the finite size of the disk is considered.
Solutions  for the cored Isothermal Sphere
are shown in Figure~\ref{fig:profile}: 
we plot the solution in terms of $r_\star/r_0$
as the core radius $r_0$ can be inferred from observations. Solutions with 
$r_\star/r_0>0.1$ are of interest as they correspond
to observable deviations in dwarf galaxies.

%We'd need $z$ to be smaller than that to get $y$ smaller. 

%%Multiple passages can cause the effect to cancel, once M_2 is moving away from M_1
%%the DM is pulled back closer to the stars. 

%%Could one look for displacements of HI gas along the line joining 2 nearby galaxies? That could %%be done statistically. 

%% If stellar disk is displaced from halo center of mass, then it must not be able to rotate in equilibrium...

%%So what about the SMassive black holes - they produce cuspy density profiles... 

%%Try 1/(r+r_c) to see the transition change.. 

\section{Dynamics within the infalling halo}
\label{sec:dynamics}

In this section we study stellar orbits within the disk of the infalling halo. We 
will highlight several observable effects of modified gravity that arise from the effect
of an external force on the structure and rotation of the finite-sized stellar disk. 
We first examine the orbital structure analytically and then show results from 
simulations of the infalling disk galaxy.

\subsection{ Dwarf galaxy parameters}
\label{sec:parameters}

We have presented above two mass profiles for dwarf galaxy halos, the
cSIS and NFW profiles. 
We will study three fiducial dwaf-galaxy halos based on these profiles,
with parameters given in Table~\ref{Table1}. 

\begin{table}[ht]
\centering
\caption{
  \protect\centering{
    Halo profile parameters for the cases studied
    below. }
}
\begin{tabular}{llccc}
  \hline
            & Eqn.  & $r_0$ (kpc) & $\rho_0$ ($\mathrm{M_\odot/kpc^3}$)\\
  \hline
  \cSISfour &(\ref{eqn:cSIS_def})  & 4 & $4 \times 10^6$    \\
  \cSIStwo  &(\ref{eqn:cSIS_def})  & 2 & $1.2 \times 10^7$  \\
  \NFWfour  &(\ref{eqn:NFW_def})   & 4 & $1 \times 10^7$    \\
  \hline
    \end{tabular}
    \label{Table1}
\end{table}

\begin{figure}[t]
\centering
\includegraphics[width=10cm]{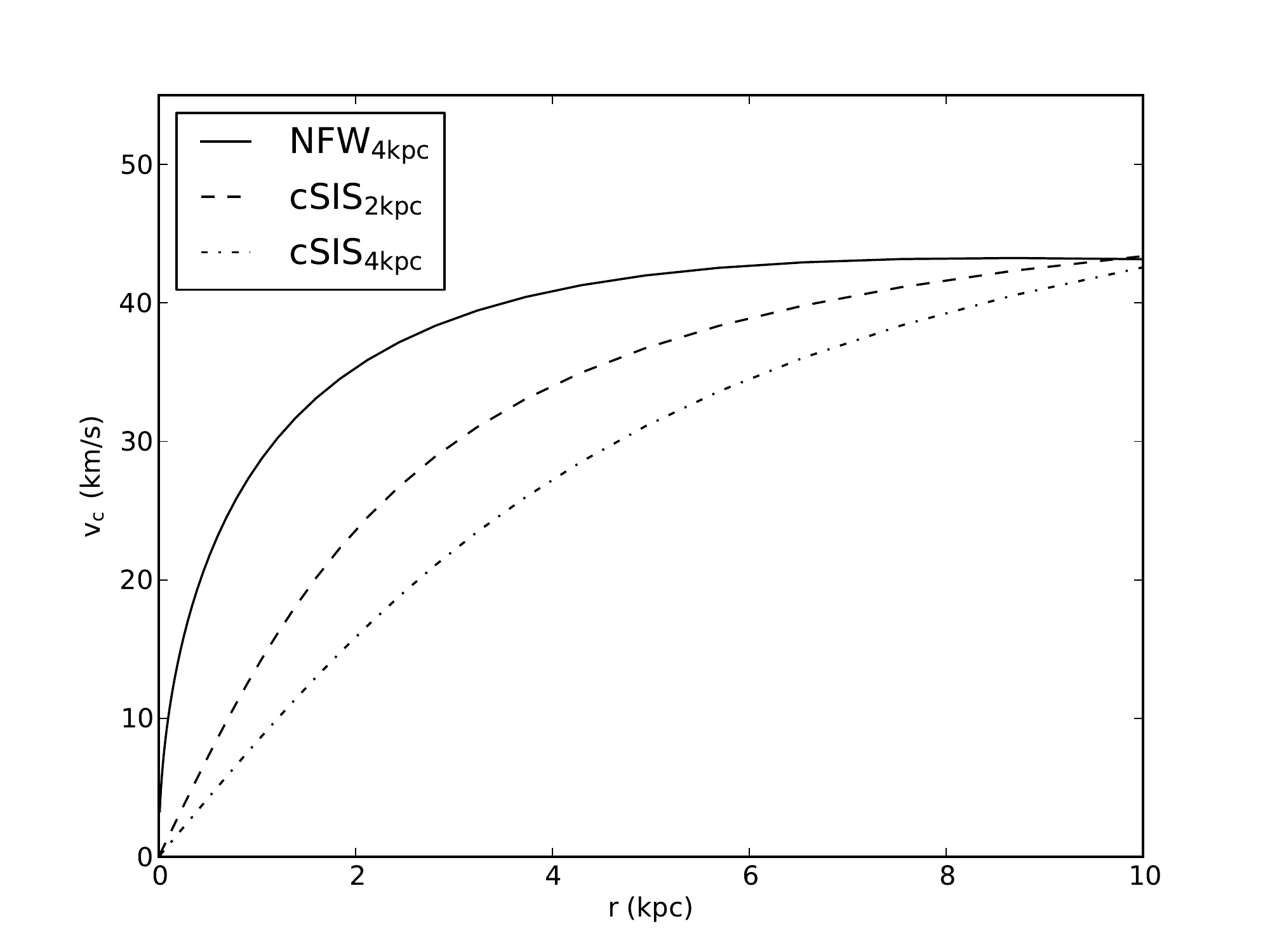}
\caption{The rotation curves for our three fiducial halos, defined in
  \S\ref{sec:parameters}:
  the cored Isothermal Sphere (cSIS) with core radius of 2 and 4 kpc,
  and the NFW profile with core radius of 4kpc.  
  The rotation curves of \NFWfour\ and \cSIStwo\ are reflective of 
  observed galaxies; \cSISfour\ is less realistic,
  but useful for studying the behavior of a disk within a flat core.
  \label{fig:rot_curves}
}
\end{figure}

\medskip
In order to study the behavior of a disk within a flat core, we will use 
the \cSISfour\ halo.  The core radius of 4kpc is on the large end for low-mass
galaxies 
\citep[e.g.][]{Swaters09}, %(e.g. Swaters et al 2009), 
but we choose it to emphasize the effects within
a flat core.  For a more realistic flat-core parameterization,
we use the \cSIStwo\ halo. For a steeper-cored case, we use the \NFWfour\
halo.  The core radius of 4kpc for an NFW profile is
well within the typical range observed in 
surveys of dwarf galaxies.  The rotation curves for these three 
fiducial cases are shown in Figure~\ref{fig:rot_curves}.

Observed late-type dwarf galaxies are rotationally supported and 
can be observed via their 21cm emission from the HI disk or 
optical data for the stellar disk. The peak rotational velocities 
are typically in the range $\sim 40-100$km/s, and the disk sizes
range from several kpc in the optical to $\sim 10$kpc in
HI. High resolution spatial and velocity information 
for galaxies at distances of $\sim 10-100$ Mpc from us
can provide
a useful sample for the effects discussed here. Some of these galaxies
are in or near groups and clusters, but there is a significant
fraction that lie in the field/void environments that could be
unscreened. More details on the observational sample 
available currently and with forthcoming surveys will be provided
elsewhere. 

We will assume in what follows that the attracting galaxy has mass 
$M_1 = 1.6\times 10^{11}M_\odot$ and lies at a distance  $d = 100\mathrm{kpc}$
from the infalling halo at the end of a period of infall, typically 
3 Gyr. For numerical results, we set $\Delta G/G_N = 1$. 
While we assume below that the fifth force on the dwarf
galaxy is due to a nearby neighbor, any unscreened or partially
screened galaxy or void can provide the force.

\subsection{Orbits in an axisymmetric potential}

As discussed in \S\ref{sec:two_body_infall}, 
in the frame of the infalling halo center of mass, 
stars feel an acceleration which the gas and dark matter
does not.  We set up a coordinate system centered on the infalling halo,
with halo 1 on the negative $z$-axis.  For a star at position 
$z$ (we had denoted it $r_\star$ above), the force in the $z$-direction due
to $M_1$ is
\begin{equation}
  \label{eqn:F_z}
  F_z = -\frac{\partial\Phi_z}{\partial z} = \frac{\Delta G  M_1}{d^2}
\end{equation}
where, for the potential, we have approximated the final two terms 
of Equation~\ref{eqn:phi_def} assuming $z\ll d$. 

We'll assume that the stars orbit within a spherically symmetric,
static background potential due to the dark matter halo.
This additional force in the $z$-direction acts to break the spherical
symmetry, resulting in an axisymmetric potential. 

\begin{figure}[t,h]
\centering
\includegraphics[width=8cm]{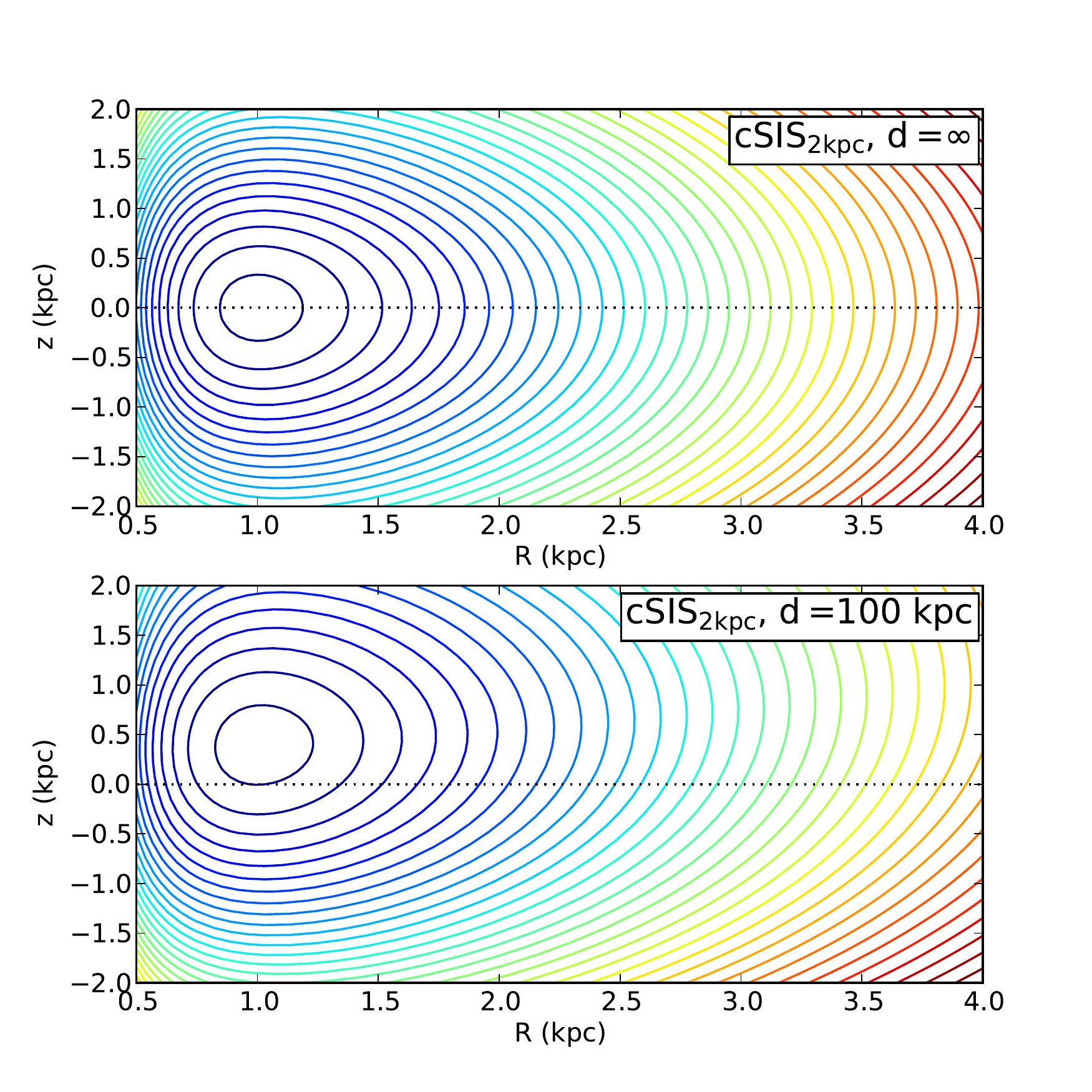}
\includegraphics[width=8cm]{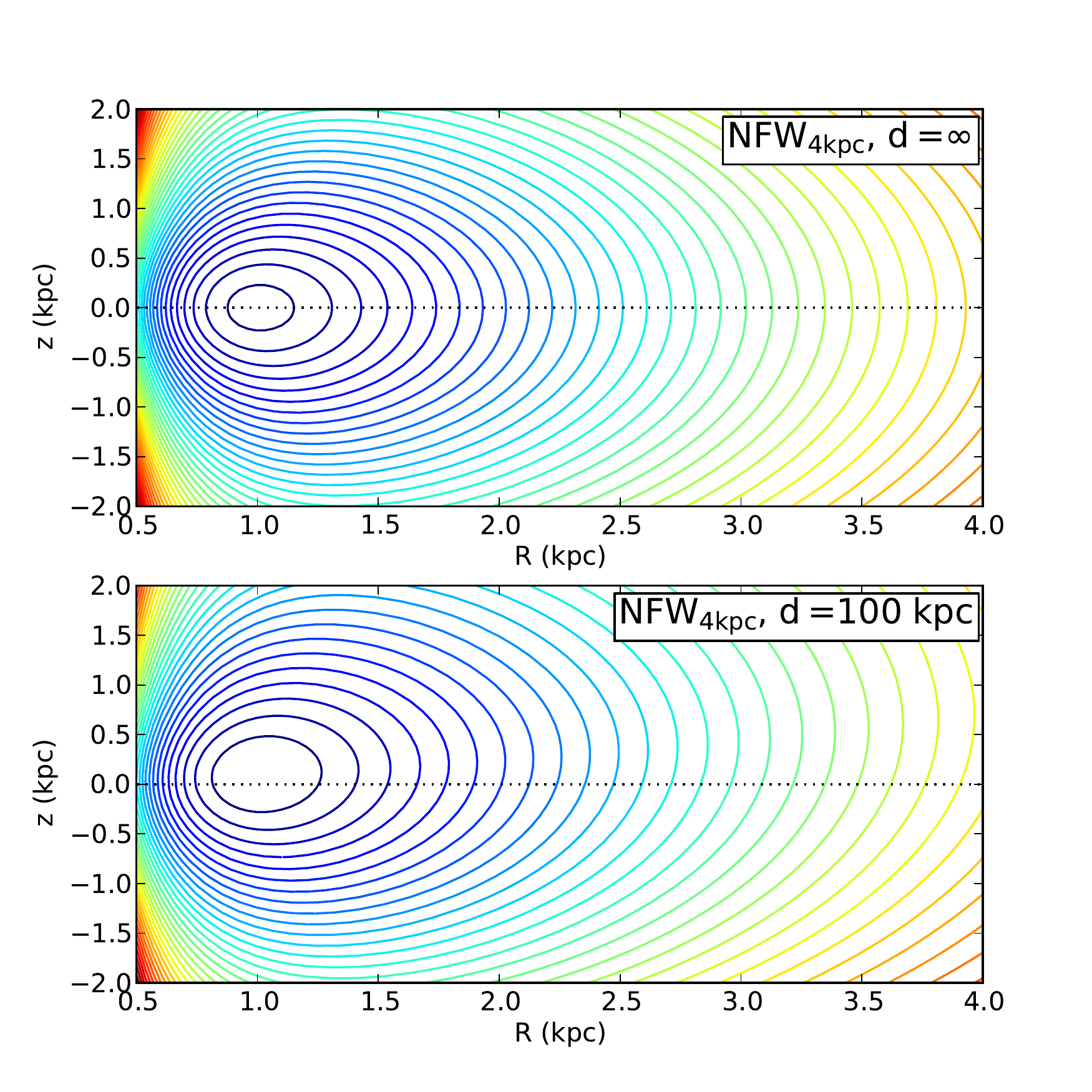}
\caption{The effective potential $\Phi_{\rm eff}$ (Eqn.~\ref{eqn:Phi_eff})
  in the meridional plane 
  for the cSIS halo with $r_0$=2kpc (left panels) and NFW halo (right panels),
  for a particle with angular momentum corresponding to
  a circular orbit at 1 kpc. The origin in each plot corresponds to the
  center of the dark matter halo. As expected for the unperturbed 
  case (upper panels), the location of the minimum is $z=0$: that is, the 
  orbits are centered on the DM halo.  The lower panels show the case 
  where the infalling halo is 100 kpc from the attracting halo.  Due to the
  self-screening of the stars, their orbits are centered above the 
  original plane of the disk. For the orbit shown here (1 kpc in radius), 
  the displacement for the cSIS halo is about 0.5 kpc, much larger than 
  for the NFW halo. 
  \label{fig:potential}
}
\end{figure}

Consider the motion of a star in a general axisymmetric potential 
$\Phi(\vec{r}) = \Phi(R,z)$. The equation of motion is given by
\begin{equation}
  \ddot{\vec{r}} = -\vec\nabla \Phi(\vec{r})
\end{equation}
where the dot denotes a time derivative. Transforming into
cylindrical coordinates $(R,\phi,z)$ with $r=\sqrt{R^2+z^2}$, 
the three components of this vector equation can be equivalently expressed
\begin{eqnarray}
  \ddot{R}-R\dot\phi^2 &=& -\frac{\partial\Phi}{\partial R} \\
  \frac{d}{dt}(R^2\dot\phi) &=& 0 \\
  \ddot{z} &=& -\frac{\partial\Phi}{\partial z}
  \label{eqn:z_motion}
\end{eqnarray}
The second of the three equations above expresses a conserved quantity: 
we define the constant of motion $L_z\equiv R^2\dot\phi$
and substitute this into the first equation to get
\begin{equation}
  \ddot{R} = -\frac{\partial}{\partial R}
  \left[\Phi + \frac{L_z^2}{2R^2}\right]
  \label{eqn:R_motion}
\end{equation}
The terms inside the brackets on the RHS now define an effective 
2D potential.  Equations~\ref{eqn:z_motion} and \ref{eqn:R_motion} completely
define the dynamics of a test particle with angular momentum $L_z$: 
the 3D motion has been reduced to motion in a 2D rotating plane.
Examples of these two dimensional effective potentials 
are shown in Figure~\ref{fig:potential}.
As the strength of the background gravitational field increases, the 
minimum of the effective potential moves above the z-axis.  Thus stars
orbiting within the frame of halo 2 will feel a force directed away from
halo 1.  In the next section, we estimate the dependence of this force 
on the distance of the star from the halo center

\subsection{Orbits in the stellar disk of the infalling halo}
\label{sec:orbits}

In the dwarf galaxy encounter described above, 
the effective potential for a stellar orbit is
\begin{equation}
  \label{eqn:Phi_eff}
  \Phi_{\rm eff}(R, z; L_z) = \Phi_r(r) - F_zz + \frac{L_z^2}{2R^2}
\end{equation}
where $\Phi_r(r)$ is the spherically symmetric potential due to 
the dark matter halo (e.g.~Eqns.~\ref{eqn:cSIS_Phi} and \ref{eqn:NFW_Phi}), 
$r=\sqrt{R^2+z^2}$, and $F_z$ is the perturbing force (Eqn.~\ref{eqn:F_z}).
The minimum of $\Phi_{\rm eff}$ defines the position $(R_0,z_0)$
of a circular orbit with angular momentum $L_z$.
Orbits which are elliptical and/or inclined will oscillate about this minimum,
so it is a good first-order approximation of the mean position of stars with
angular momentum $L_z$. 
Setting the partial derivatives of $\Phi_{\rm eff}$
with respect to $z$ and $R$ to zero gives
\begin{eqnarray}
  \frac{1}{r_0}\left.\frac{d\Phi_r}{d r}\right|_{r_0} 
  &=& \frac{L_z^2}{R_0^4} \nonumber\\
  \frac{1}{r_0}\left.\frac{d\Phi_r}{d r}\right|_{r_0} 
  &=& \frac{F_z}{z_0}\label{eqn:Rz_constraints}
\end{eqnarray}
Given an expression for the halo potential $\Phi_r(r)$,
these equations can be solved for the position $(R_0,z_0)$
of a star with angular momentum $L_z$.
Let's first explore the simple case of a star in a circular orbit within the
potential.  A circular orbit has $L_z^2 = (R_0v)^2 \approx R_0G_N M_2(<R_0)$,
where the final approximation holds for $z_0 \ll R_0$.  Inserting this into
Equations~\ref{eqn:Rz_constraints} gives,
\begin{equation}
  z_0 \approx \frac{F_z R_0^3}{G_N M_2(<R_0)}
\end{equation}
This shows that for any realistic halo, where $M_2(<R_0) \propto R_0^\alpha$
with $\alpha < 3$, the mean offset should increase as a function of
radius.  The precise nature of this increase will depend on the exact 
shape of the potential.

\begin{figure}[t]
\centering
\includegraphics[width=8cm]{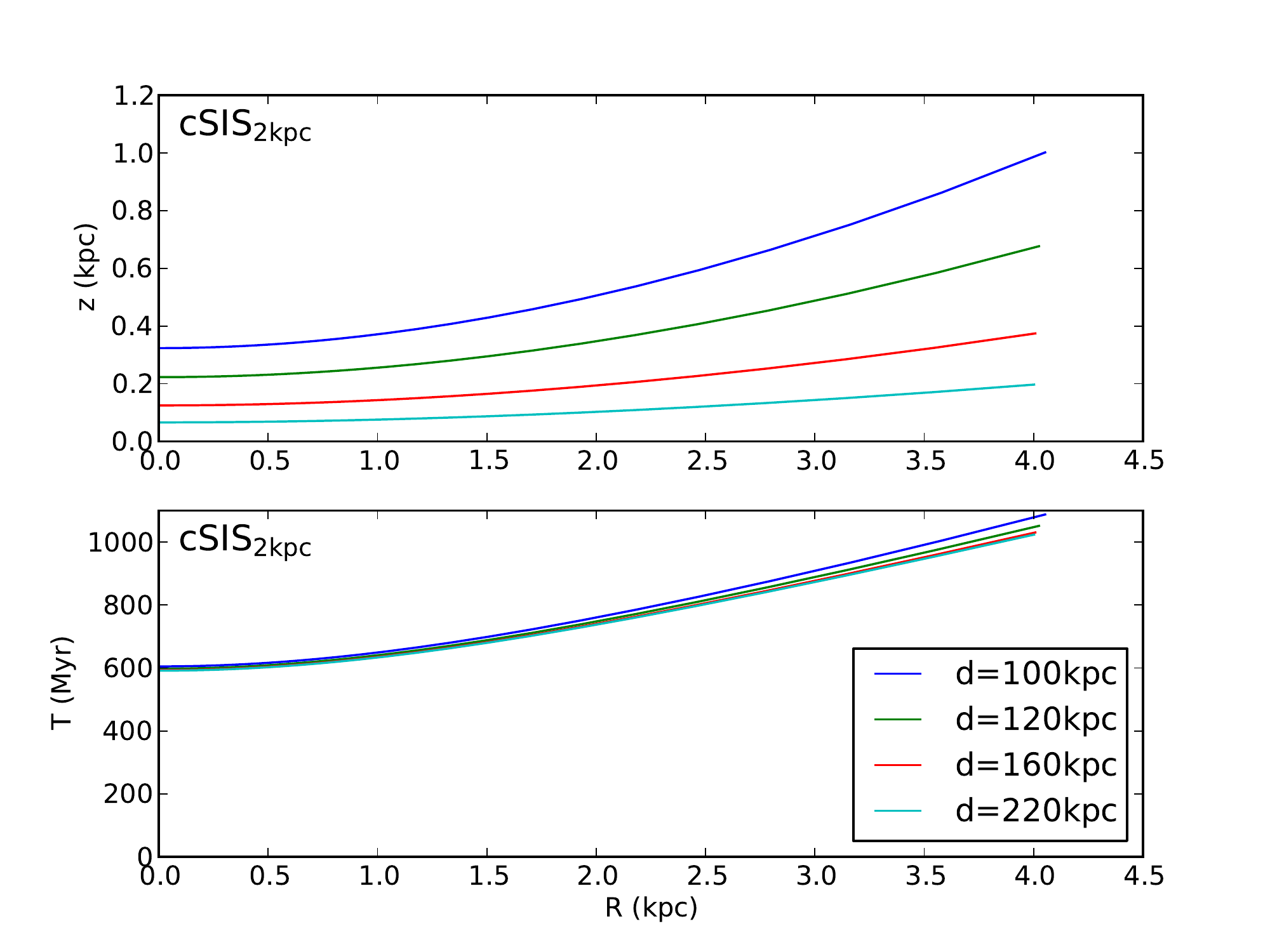}
\includegraphics[width=8cm]{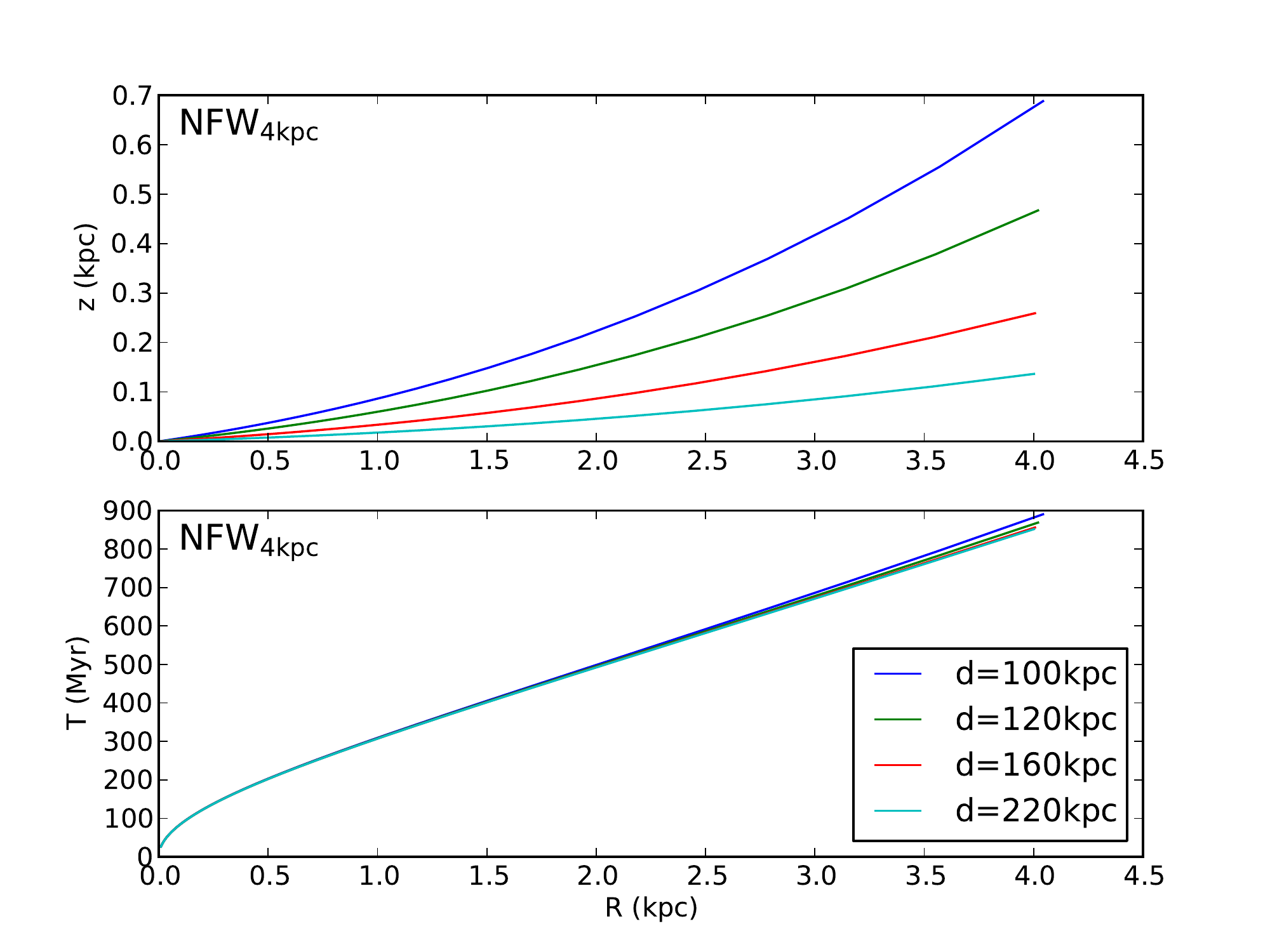}
\caption{
  \label{fig:eff_min}
  Top panels show the numerically determined 
  relationship between $z_0$ and $R_0$
  (Eqn.~\ref{eqn:Rz_constraints}) for the cSIS halo with $r_0=2$kpc (left), 
  and NFW halo (right).
  As the distance $d$ decreases, the magnitude of the vertical offset increases
  as $d^{-2}$.  As expected from the discussion in \S\ref{sec:orbits}, 
  the offset of the circular radius increases with $R_0$, and increases
  more steeply for the NFW case.
  Bottom panels show the vertical oscillation period as a function of $R_0$
  (see \S\ref{sec:oscillations}) calculated via 
  Equations~\ref{eqn:frequency}-\ref{eqn:T_z}.
}
\end{figure}

\subsection{Vertical oscillations}
\label{sec:oscillations}
In the limit that the $z$-direction potential is turned on very slowly,
the stellar orbits would be expected to adiabatically trace 
the minimum of the effective potential.  
Outside this limit, the perturbation of the effective potential
will cause the disk to oscillate coherently 
in the $z$-direction.  We can understand this
oscillation by approximating the vertical shape of the potential to 
second order near the minimum.  Then the vertical oscillation frequency can
be approximated by that of a simple harmonic oscillator 
\begin{equation}
  \label{eqn:frequency}
  \omega_z^2(r_0) = \left.\frac{1}{2}
  \frac{\partial^2\Phi_{\rm eff}}{\partial z^2}\right|_{r=r_0}
\end{equation}
From Equation~\ref{eqn:Phi_eff} we can write
\begin{eqnarray}
  \label{eqn:d2Phi_dr2}
  \frac{\partial^2\Phi_{\rm eff}}{\partial z^2} &=& 
  \frac{1}{r}\left(1-\frac{z^2}{r^2}\right)\frac{d\Phi_r}{dr}
  + \frac{z^2}{r^2}\frac{d^2\Phi_r}{dr^2}\nonumber\\
  &=& \frac{G_N M_2(<r)}{r^3}\left(1-3\frac{z^2}{r^2}\right)
  + 4\pi G_N \rho(r)\frac{z^2}{r^2}
\end{eqnarray}
The second line follows from Poisson's equation and geometrical
considerations.
Equations~\ref{eqn:Rz_constraints} and 
\ref{eqn:frequency}-\ref{eqn:d2Phi_dr2} can then be used to understand the 
frequency of radial oscillations within a particular potential.  
The qualitative behavior can be seen in the limit $z/r \ll 1$.
In this case the period of the oscillations is
\begin{equation}
  \label{eqn:T_z}
  T_z \equiv \frac{2\pi}{\omega_z} \sim \left[\frac{R_0^3}{M(<R_0)}\right]^{1/2}
\end{equation}
As noted above, any realistic halo will be bounded by $M(<R_0)\sim R_0^\alpha$ 
with $\alpha < 3$.  This leads generally to an oscillation period 
that increases with radius.  

Thus for steep profiles, when the potential is turned on non-adiabatically,
we expect the perturbations to cause
vertical oscillations about the minimum of the effective potential
with a radial phase dependence: a disk with initially near-circular 
orbits will display vertical waves which propagate radially 
outward with time.

As seen from the lower panels of Figure~\ref{fig:eff_min}, for the outskirts
of the disks in both the cSIS and NFW halos,
the infall time of 3Gyr is only $3-4\times$ the oscillation period: 
thus we expect the outer parts of the disk to display coherent 
vertical oscillations.
In the steep inner-core of the NFW halo, the oscillation
timescale is much smaller, meaning that we expect the coherence to
be washed out due to orbital scatter within the disk.

\begin{figure}[t,h]
\centering
\plotone{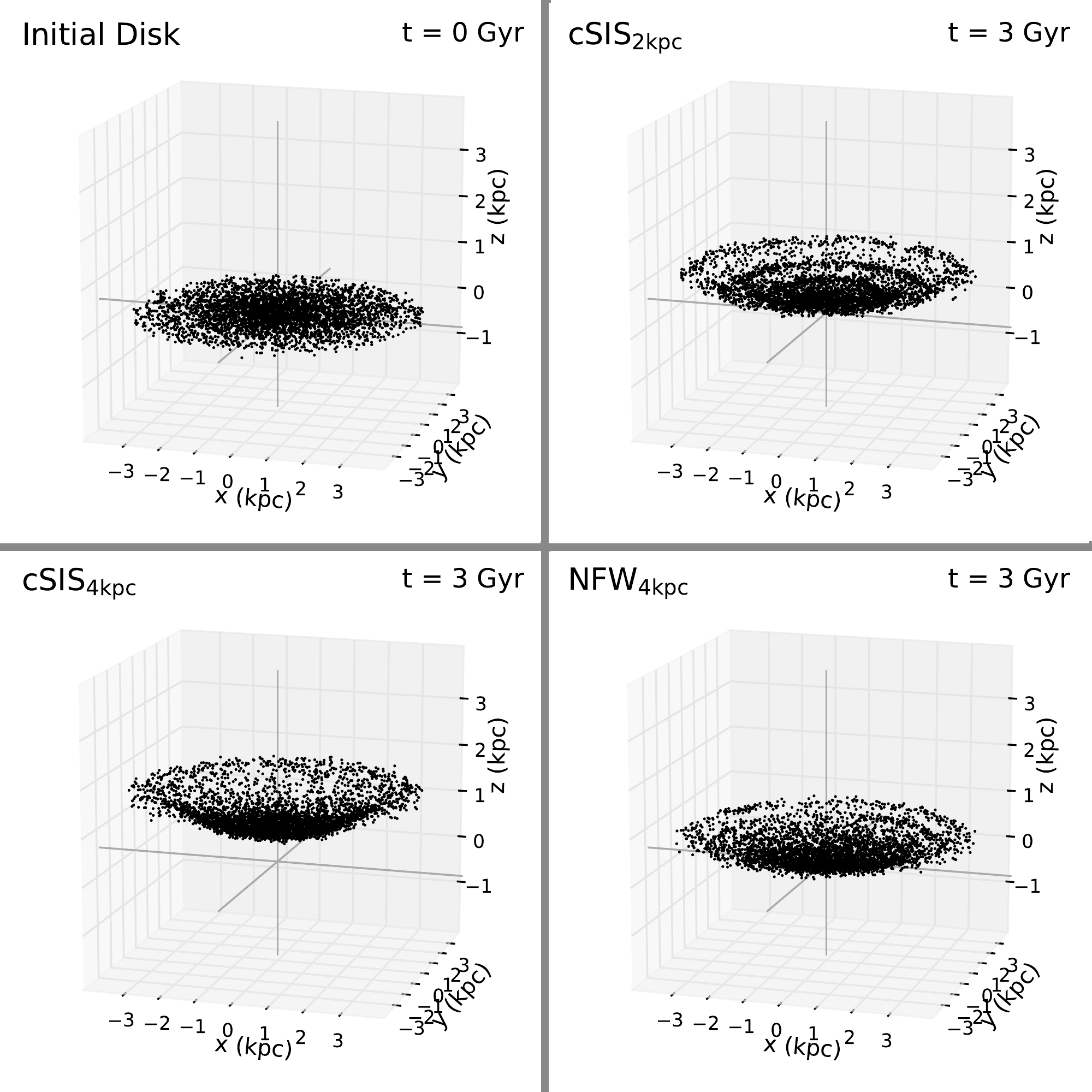}
\caption{The change in the morphology of stellar disks in infalling
  dwarf galaxies is shown for a face-on orientation with respect to
  the neighboring galaxy.  The initial disk is shown in the upper-left
  panel.  The final  disks for the \cSISfour, \cSIStwo, and
  \NFWfour\ halos are shown in the remaining panels for 
  infall starting from a distance of 240kpc and ending 3Gyr later at a 
  distance of 100kpc from the larger neighbor galaxy. 
  The neighbor  lies on the negative $z$-axis.  Each dot shows the 
  position of a star taken to be initially on a near-circular orbit. 
  The radially averaged profiles for
  the face-on case are shown in Figure~\ref{fig:final_profile}, and
  the rotation curves for the edge-on case are shown in 
  Figure~\ref{fig:rot_curve_comp}.
  \label{fig:final_disk:faceon}
}
\end{figure}

\begin{figure}[t,h]
\centering
\plotone{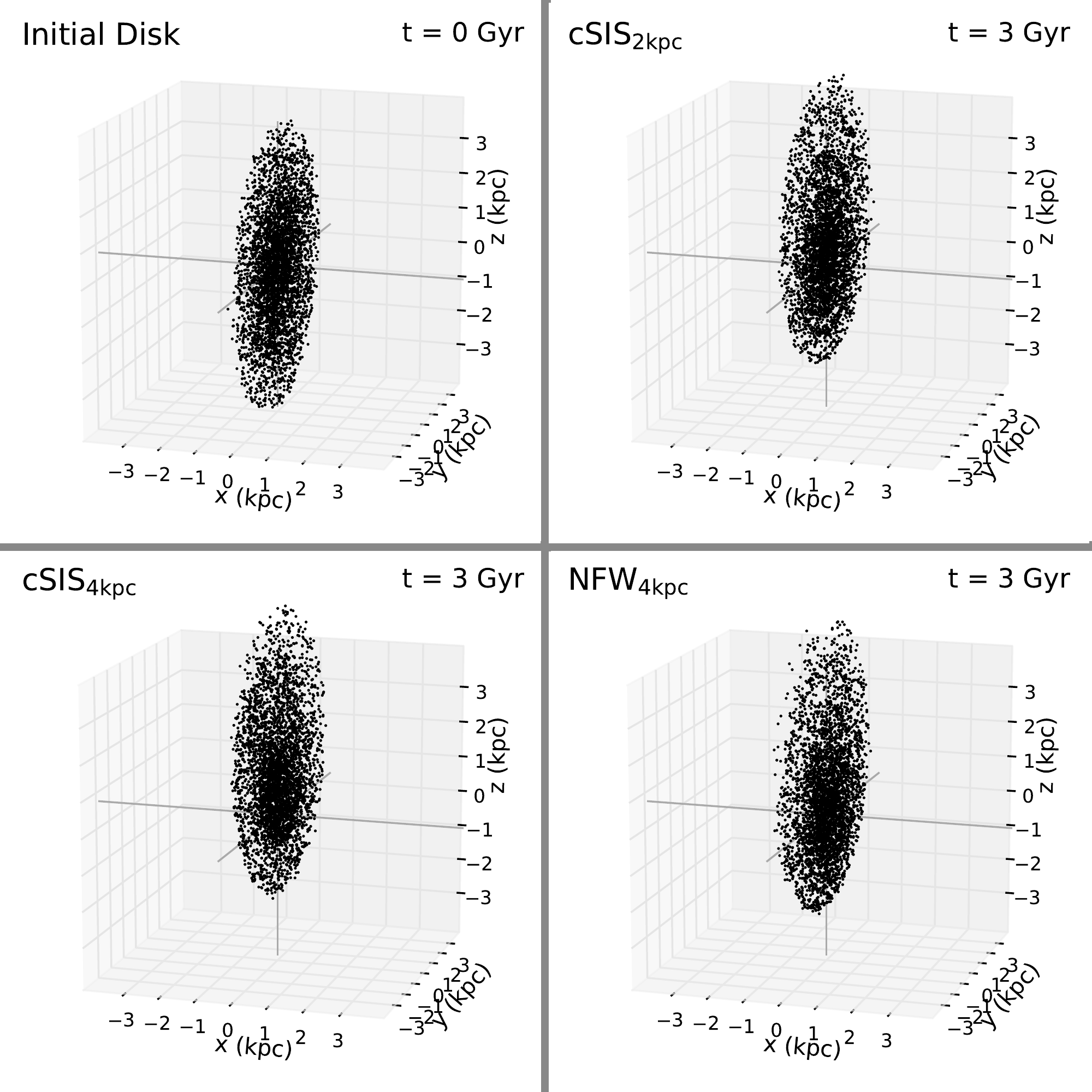}
\caption{The change in the morphology of stellar disks in infalling
  dwarf galaxies is shown for an edge-on orientation with respect to
  the neighboring galaxy.  For details refer to the caption of 
  Figure~\ref{fig:final_disk:faceon}.
  \label{fig:final_disk:edgeon}
}
\end{figure}

\begin{figure}[t,h]
\centering
\includegraphics[width=16cm]{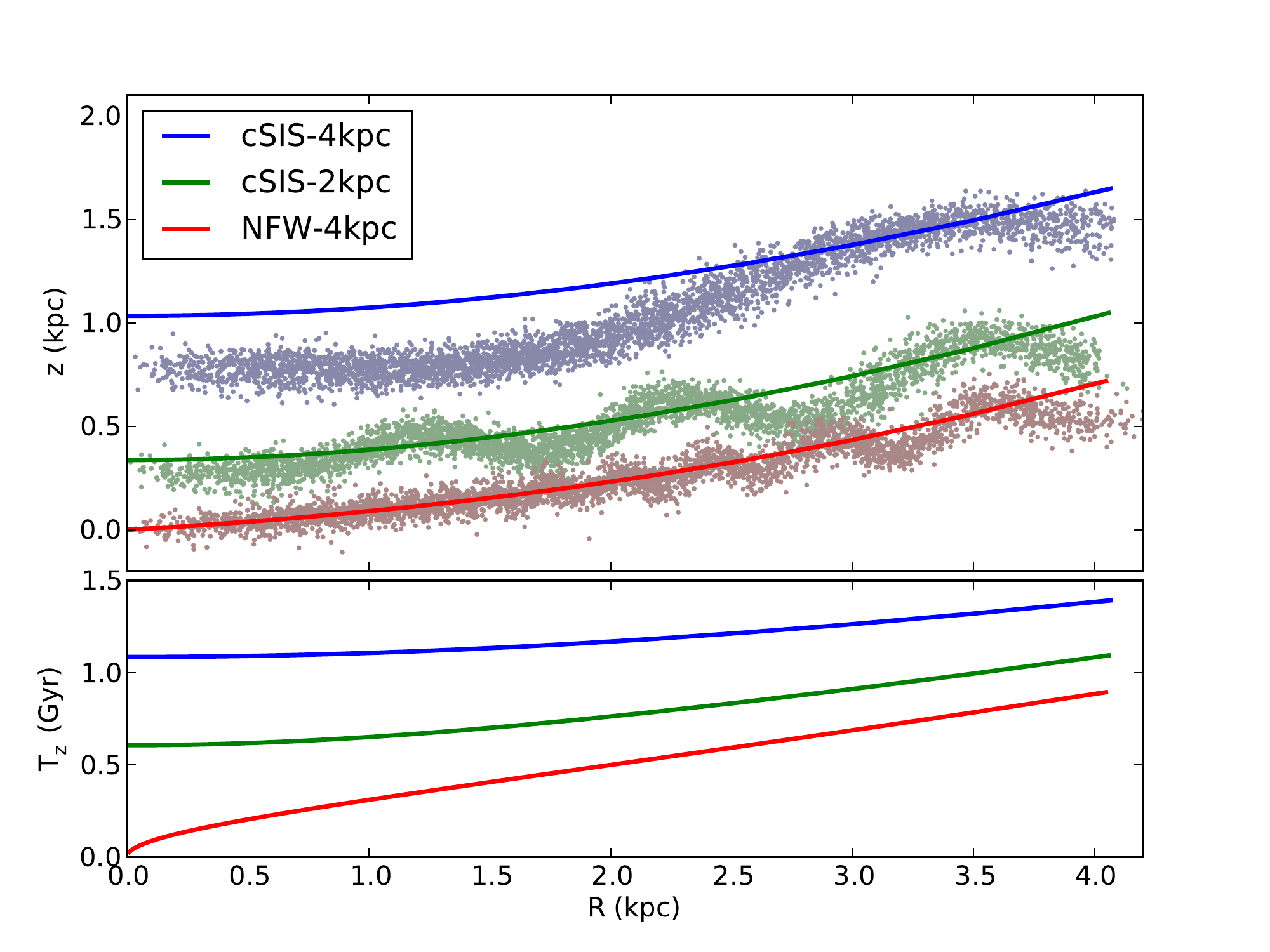}
\caption{
 {\it Upper panel}: The distribution of the height of stars $z$ at 
 different radii $R$ in the face-on case (Fig.~\ref{fig:final_disk:faceon}).  
 The lines denote the minimum of the effective
 potential for stars of increasing $R$ (and angular momentum).  The 
 stars don't lie precisely at this minimum because their orbits
 are generally inclined, meaning they oscillate vertically about the minimum
 of the effective potential. {\it Lower panel}: the oscillation period 
 $T_z$ as a function of radius.  Because the
 integration lasts 3Gyr, we'd expect the number of oscillations to be roughly
 equal to $n = 3/(T_z/Gyr)$.  For the NFW case, $T_z$ increases rapidly with
 radius, which leads to the narrow oscillations in the disk.  
 \label{fig:final_profile}
}
\end{figure}

\begin{figure}[t,h]
\centering
\includegraphics[width=12cm]{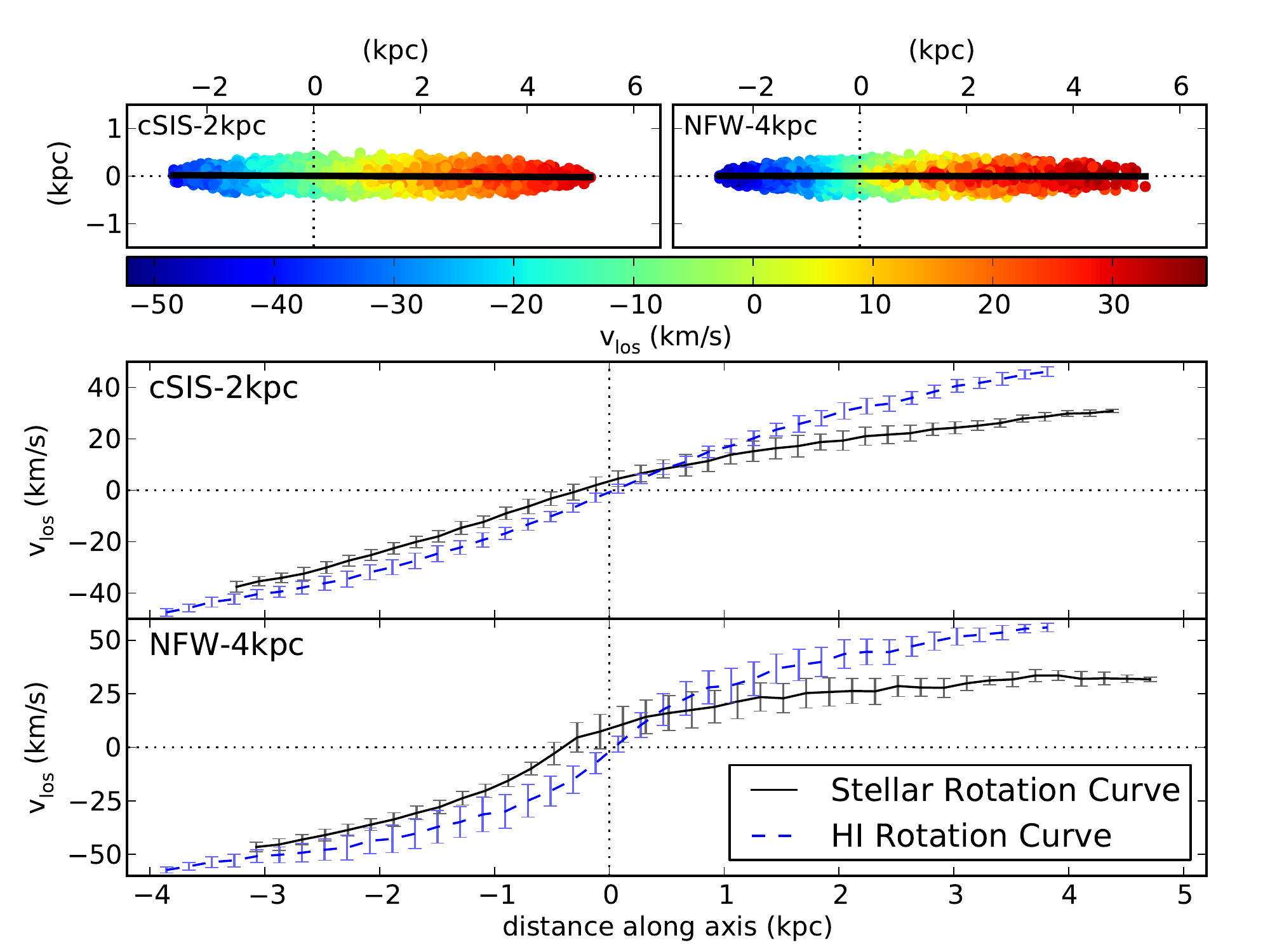}
\caption{
  {\it Upper panels}: A near edge-on view of the final disk from the edge-on
  infall case (Fig.~\ref{fig:final_disk:edgeon}).  We show only the \NFWfour\
  halo and the \cSIStwo\ halo here; 
  the rotation curve of \cSISfour\
  is qualitatively similar to that of \cSIStwo.
  Stars are color-coded by their radial
  velocity in km/s.  Black lines denote the principal axis of the stellar
  distribution used to define the rotation curves.
  {\it Lower panels}: The HI and stellar rotation curves along the major 
  axis of each galaxy.  Two effects are apparent. First, because the HI 
  disk is not dense 
  enough to be self-shielded, it feels a stronger gravitational attraction to
  the dark matter and has a faster overall rotation.  
  Second, the self-screened stellar disk 
  lags the halo, leading to asymmetric distortions in the
  rotation curve of the disk.  This effect is more pronounced for 
  a steeper core.
  \label{fig:rot_curve_comp}
}
\end{figure}

\subsection{Simulations of the stellar disk}

To see all of these dynamical effects in action, we simulate the orbits
of 4000 stars in a 4kpc exponential disk within the infalling halo.
We allow the halo to fall from an initial distance
$d=240$kpc to a final distance of $d=100$kpc in 3Gyr, so that the final
infall speed is approximately 100km/s.  The initial orbits in the disk are near circular, 
with a Gaussian scatter of $\sim$1 km/s in each direction.
We integrate the orbits using the 
\texttt{scipy} interface to the \texttt{LSODA} routine in the 
\texttt{ODEPACK} library.\footnote{\url{http://www.scipy.org}}  
We make a few approximations consistent with our assumptions about
dwarf galaxies in \S\ref{sec:two_body_infall}: 
first, we assume the DM in the infalling halo gives rise to a static background
potential, that is, the shape of the halo does not react to the changing
position of the stellar disk or the tidal effects from the distant halo.
Second, we assume that the dominant force on the stellar disk is from the
dark matter: we ignore self-gravitation of the disk itself.

For each halo, we try two limiting configurations: an edge-on infall and a
face-on infall, as seen from the attracting galaxy.  The results are shown in
Figures~\ref{fig:final_disk:faceon}-\ref{fig:final_disk:edgeon}.  
They lead to the following distinct observable effects (discussed further in the
next section).

%{\bf Summary of observable effects of modified gravity dwarf galaxy disks:}
\begin{enumerate}
\item For both the edge-on and face-on cases,
  the mean position of the stellar disk is displaced from that of its DM
  halo.  As expected from the discussion in \S\ref{sec:two_body_infall}, this
  displacement is on the order of $\sim 1$kpc for the cSIS halo, and is
  less pronounced for the NFW halo.  Though the location of the dark matter
  halo is difficult to determine observationally, in this case the gas and
  dust in the galaxy is expected to remain unscreened, and thus track the
  location of the dark matter halo.  This leads us to expect a
  separation of  up to $\sim 1$kpc between the stellar disk and HI disk of
  a suitable dwarf galaxy.
  
\item The face-on infall cases show distinct
  morphological features: the galactic disks become ``cup-shaped'', with the
  outer regions of the disk displaced further than the inner regions,
  as expected from the discussion in \S\ref{sec:orbits}.  
  Figure~\ref{fig:final_profile} shows an axially projected view of the final
  location of the stellar disk.  The precise shape of this distortion 
  will depend in a predictable way on two factors: 
  the inner slope of the DM halo density profile, which can be observationally
  constrained from the rotation curve, and the orientation of the disk
  relative to the background potential, which can be deduced from dynamical
  surveys of nearby galaxies. The edge-on infall case shows asymmetric 
  tidal features -- the near side of the disk (relative to its neighbor) 
  is more compact than the far side. 
  This feature should also be kpc-scale in magnitude. 

\item The asymmetry in the disk for edge-on infall is
 reflected in the observable rotation curves of the stellar
  disk (see Fig.~\ref{fig:rot_curve_comp}).  Again, this induced asymmetry in the
  disk rotation is expected to depend in predictable ways on the halo profile
  and orientation with respect to surrounding structure: in particular, the
  rotation curve is more extended on the far side of the disk.
    
\item Finally, because we generally expect the gaseous components of the
  galaxy to remain unscreened, the observed rotation of the HI disk will
  be enhanced by a factor of $\sqrt{1+\Delta G/G_N}$ relative to that of
  the stellar disk.  The magnitude of this enhancement depends on the
  parameters of the MG theory, but may be as much as 20-40\% as shown 
  in Figure~\ref{fig:rot_curve_comp}.
  This effect should be associated with the morphological
  and dynamical effects mentioned above.
\end{enumerate}

\section{Discussion}
\label{sec:discussion}

\subsection{Summary of results}

We have presented four different observable effects of 
gravity in MG theories on small disk galaxies: 
displacement of the stellar disk relative to the center of mass of 
the DM and gas, cup-shaped distortions in the stellar disk, asymmetries in
the stellar rotation curve, and differences in rotational velocity between
stellar and HI disks.

The offset and distortion of the stellar disk (\#1 and \#2, above),
which are up to $\sim$1kpc, are potentially observable in nearby dwarf galaxies, 
though care must be taken to rule out non-MG effects.
The displacement of the stellar and HI disks is the most promising
of the morphological effects
(though the resolution with which the HI disks are typically imaged may restrict
this to galaxies within $\sim 100$Mpc). In GR, such displacements due to 
ram-pressure stripping of the gas are predicted in clusters 
\citep{Farouki80,Vollmer03}, %(Farouki \& Shapiro 1980; Vollmer 2003), 
but not in voids where MG becomes important.  Moreover the signal has
a clear correlation with the direction of the gravitational field. 
%Such a separation in a low-density environment
%would be difficult to explain within the context of GR.

The combination of the two rotation curve effects (\#3 and \#4 above)
provides the strongest observable test, as it corresponds to discrepancies of 
about 10 km/s between the HI and stellar rotation curve. 
The discrepancy has a clear signature, as shown in 
Figure~\ref{fig:rot_curve_comp}. 
As discussed above, the distortions shown by a particular disk 
galaxy are sensitive to its inner halo structure. 
Since the HI rotation curve does tell us about the 
inner mass profile, one can map the observed rotation curve 
to the expected deviation in the stellar disk. 
In addition, the signal can be distinguished from contamination
from other astrophysical effects by its dependence on environment
and orientation: with a sufficiently large sample of late-type dwarf
galaxies, one could obtain constraints that are robust to the 
varied dynamical histories of the observed galaxies. The signal-to-noise can be 
improved with a  sample of galaxies 
that are expected to be unscreened.  

The magnitude of these predicted effects will depend on various parameters 
of the disk galaxy and the theory: 
1. The morphological effects scale as $(1 + \Delta G/G_N)$, 
which is taken to equal 2 above,
while the rotation curve effects have a weaker dependence as they 
scale as the square root of the same factor. 2. The shape of the rotation curve
determines the overall offset of the stellar disk and the warping of its outer
edge. 3. The observable asymmetry in the morphology and rotation 
curve depends on the orientation of the galaxy disk relative to its 
neighbor and its inclination relative to us. 
4. Predicted deviations depend on the ratio of halo masses and the 
separation $d$ of the galaxy 
centers. However, a nearby neighbor is not essential -- 
in general the external force on a galaxy is determined by 
its complete environment.

\subsection{Comparison with previous work}

We introduced the work presented above in \S\ref{sec:introduction} 
by noting the connections to 
astrophysical tests of the fifth force and the equivalent principle. 
Those tests were discussed by 
Gradwohl \& Frieman (GF) (1992), \nocite{Gradwohl92}
Kesden \& Kamionkowski (KK) (2006) \nocite{Kesden06}
and references therein. 
These authors placed limits on additional 
interactions between the dark matter particles that are not 
felt by baryons. GF 
considered the Milky Way (MW) infall towards Andromeda (M31), while KK 
examined the leading and trailing tidal arms of dwarf galaxies in 
the Milky Way. 
For related effects on large-scales, see also \citet{Keselman10}.

The motivation for our work is different: we are considering tests of modified 
gravity theories that introduce an additional, universal interaction for all
matter -- one that is suppressed in massive galaxies like the Milky Way but 
may act on smooth dark matter and gas in smaller galaxies in low-density 
environments. Hence the tests within the MW and M31 considered by previous 
authors would not be the most powerful tests of modified gravity. 

But the physical effects in dwarf galaxies described in this paper 
would qualitatively apply to the MW and M31 stellar disks. The 
absence of the distortions of the stellar disks of the MW and M31 
and the lack of asymmetry in rotation curves places some constraints 
on the dark matter-only fifth force (see GF for a detailed discussion). 
The MW is oriented edge-on towards M31, while M31 is partially face-on 
\citep[e.g.][]{Cox08}, %(e.g. Cox \& Loeb 2008), 
so many of the observable effects we discuss 
could occur. However we have not attempted 
to quantify the constraints on the DM-only interaction from the 
regularity of the disks of the MW and M31.  

\subsection{Constraints on gravity theories}

The three primary screening mechanisms in the literature are chameleon, 
symmetron and Vainshtein screening. In Appendix A we summarize their 
properties and note that for astrophysical tests of the kind discussed 
above, chameleon and symmetron theories would show observable signatures. 
Vainshtein screening protects the interior of the halos of all masses, 
so it is not likely to show any deviations in the stellar disks (but see discussion
in Appendix A). 

For observational tests on small scales, as discussed in Appendix A, 
chameleon theories may be approximately represented by
 the two parameters: the coupling constant $\hat \alpha \approx \Delta G/G_N$ 
 and the value of the Newtonian
 potential $\Phi_N^{\rm tran}$ that provides the transition from screened to un-screened halos. 
 The dwarf galaxy tests discussed here can set upper limits on $|\Phi_N^{\rm tran}|$ for 
 given $\hat\alpha$. 
 
The fact that stellar disks in the literature do not consistently show any of the deviations 
described above can place constraints on chameleon theories 
(provided $\hat\alpha$ is not much smaller than unity). 
Published measurements 
of late-type dwarf galaxy images in the optical and HI as well as rotation curves 
based on stellar and HI data appear to have the resolution to detect the signatures
discussed above 
\citep[e.g.][]{Swaters09}. %(e.g. Swatters et al 2009). 
A sample of up to a hundred useful galaxies is 
available, which allows for tests of screened vs. unscreened dwarf galaxies
 with $|\Phi_N| \lsim 10^{-7}$. Even a quick look at the literature rules out
 theories with $\hat\alpha$ significantly larger than unity. 

The constraints that can be obtained from dwarf galaxy observations 
are potentially three to four orders of magnitude stronger 
than the best astrophysical tests in the literature 
(e.g.~Schmidt et al 2009 and Lombriser et al 2010 constrain the 
equivalent parameter of $f(R)$ theories to be larger than about $10^{-4}$).  
\nocite{Schmidt09,Lombriser10}
This comes with several caveats: the 
role of astrophysical effects which impact small scales, the
robustness of our predictions  
(which needs verification with detailed simulations), and the translation
from screened dwarf galaxy halos to the parameters of the background cosmology
which has a weak model dependence (see \S\ref{sec:remaining_uncertainties} 
below). 
Nevertheless, observations such as the ones discussed above can constrain the 
parameter space of  $\hat\alpha$ and $\Phi_N^{\rm tran}$. 
A more detailed study of constraints available from current
observations will be presented elsewhere.  

Finally we note that the tests discussed here apply to a dynamical dark energy
scenario as well \citep{Khoury04}: they constrain a possible coupling of the dark 
energy to matter, and stringent constraints would point to the existence
of a new symmetry that forbids the two fields from interacting. 
In fact chameleon theories are already constrained by solar
system tests  to have a Compton wavelength close to Mpc-scales 
\citep[][J. Khoury, private communication]{Faulkner07}. 
This may limit the appeal of such theories, but as far as 
astrophysical tests go, it implies that large-scale structure tests
will not be able to detect any 
deviations: interactions on scales larger than  $\sim 10$ Mpc will not
be affected by the scalar force.  
However the dwarf galaxy tests discussed above could still test fifth
force effects. 

\subsection{Remaining uncertainties and further work}
\label{sec:remaining_uncertainties}
This study is intended to be an initial exploration into the observable
effects of general scalar-tensor gravity theories on dwarf galaxies 
moving within an external gravitational potential.  As such, a number
of questions remain.

We have relied on some simplifications in treating the orbital dynamics: 
in particular the self-gravity of the disk, and the interaction of the stellar
and HI disks have been neglected. Ignoring the self-gravity of the stellar 
disk can change the results if the galaxy is not 
dark matter dominated on scales of the observed disk. 
Dwarf galaxies generically do satisfy our assumption of being dark matter 
dominated, but a self-consistent simulation is in order. The interaction of
the stellar and HI disks is a more interesting effect to consider: we estimate 
that the stellar disk has sufficient surface density to partially displace the HI disk, 
but this depends on disk radius. So while the separation of the two disks 
may be lower than our estimates at small radii, this can lead to morphological
changes to the HI disk in unscreened galaxies -- another potentially 
observable effect. 

We have studied only a limited range of infall scenarios. 
Though a useful approximation, galaxy encounters are rarely head-on.
In addition, the ``edge-on'' and ``face-on'' situations probe only the
extremes of the possible geometries.  The majority of observed galaxy
encounters will lie between these two extremes, leading to effects in
both morphology and velocity profiles. We have checked that the
effects we describe are evident in infall with finite impact
parameters as well. 
Detailed simulations that capture a wider range of infall and
merger geometries would be useful. 

In disk galaxies there are several potential astrophysical sources of 
differences in the inferred dynamics of the HI and stellar disk (gas pressure 
on the HI disk, scattering by a bar for the stellar disk and so on). 
Any interpretation of observed galaxies will need to carefully account 
for these.  One useful observation in correcting for these effects is
that they are not likely to  correlate with either the direction of infall or
the background potential.  Thus with a large enough sample
of dwarf galaxies, the effects due to modified gravity 
could be separated from other astrophysical effects 
in a robust way. A sample of hundreds of late-type dwarf galaxies, split 
into a ``control'' sub-sample of galaxies that are expected to be screened and 
another sub-sample that are unscreened, should provide powerful tests of gravity. 

We also note that the direction of the effective force on the 
stellar disk is determined not just by its nearest large neighbor,
but by the ``local'' gravitational environment: this can include 
voids, filaments and any large but unscreened object \citep{Zhao11}.
A relatively detailed study of 
the mass distribution in the local universe, coupled with simulations 
of nonlinear screening, is needed to determine the criterion for the 
force on any
particular galaxy. On the other hand the sample of field dwarf galaxies is
much larger if one is not restricted to having a neighboring galaxy with 
the right properties. 

The level of screening of a galaxy can be 
a subtle and model-dependent question. In particular, inferences from 
spherically symmetric models can be misleading in certain environments. 
A careful mapping of simulations to the local universe can help build samples
of unscreened dwarf 
galaxies. Indeed, as noted in Appendix
A, it is possible that theories that deploy Vainshtein screening can
also be tested if galaxies lie in strong enough tidal fields that the
expectation of screening based on a spherical model is violated. 
Another possibility is to be 
theory-agnostic and simply use a set of observational criteria that 
characterize the environment of the test galaxy. Observed morphological
and dynamical deviations can be correlated with the environment to 
search for signatures of modified gravity. 

In addition to tests with late-type dwarf galaxies considered here, 
other tests of gravity are possible with low-$z$ galaxies. 
Dwarf ellipticals/spheroidals 
may exhibit asymmetries with a similar physical origin. They are known to have
very regular structure and dynamics, which can set constraints on gravity 
theories. The infall
of galaxies onto massive systems like groups and clusters is a potential test, 
especially of theories that rely on Vainshtein screening - see 
\citet{Hui09}, %Hui, Nicholis and Stubbs (2009), 
\citet{Schmidt10}, %Schmidt (2010) 
and \citet{Jain11} %Jain (2011) 
for discussions. Observable properties
of stars may also be altered  by the enhanced 
forces in MG theories \citep{Chang11,Davis11}. 
Additionally, if stars forming in dense clouds become partially screened,
it may alter the observable properties of the resulting stellar population
(M. Kesden, private communication).

\section{Conclusion}
\label{sec:conclusion}

We have presented a first study of the potential observable effects of
modified gravity   on the morphology and dynamics of
dwarf galaxies in the nearby universe. 
We consider the class of scalar-tensor theories that rely on chameleon 
or symmetron screening to recover GR in the Milky Way -- these  include all $f(R)$ models. 
We estimate the
magnitude of four potentially observable effects:
\begin{itemize}
  \item The spatial separation of the stellar disk from the dark matter and gas
  \item Distinct morphological and dynamical effects in the stellar disk
    \item An asymmetry  in the rotation velocity curve of the stellar disks along the
    direction of infall.
  \item Differences  in the rotational velocities of the stellar and gaseous disks.
\end{itemize}

Through both analytical arguments and simple orbital simulations, we show that
these effects are potentially observable in nearby dwarf galaxies ($z < 0.1$).
In particular, we estimate the magnitude of the spatial separation and
morphological distortions to be up to $\sim 1$kpc, and the
effect on rotational velocities to be  $\sim 10$ km/s. 
Though accounting for other potential astrophysical sources of 
these effects in an isolated case is nontrivial, one can exploit the
dependence on environment and orientation in order to obtain constraints on
MG with a sample of galaxies. 

Dwarf galaxies have Newtonian potentials in the range $\Phi_N\sim 10^{-8}-10^{-7}$.  
Coupled with the large force enhancements that
are generically expected in MG theories on small scales, dwarf galaxies 
have the potential to  constrain certain regions of parameter space more stringently than 
 cosmological-scale measurements and even  solar system tests.  

{\it Acknowledgements:} We are grateful to Lam Hui, Mike Kesden, Justin Khoury and especially
Mike Jarvis for several helpful suggestions. We benefited from stimulating discussions with 
Gary Bernstein, Mariangela Bernardi, Anna Cabre, Joseph Clampitt, Andy Connolly, Neal Dalal, 
Julianne Dalcanton, Kurt Hinterbichler,  Kazuya Koyama, 
Fabian Schmidt, Masahiro Takada, Mark Trodden, Vinu Vikraman, Matt Walker 
and Peter Yoachim. This work is supported in part by NSF grant AST-0908027 
and DOE grant DE-FG02-95ER40893.

\bibliography{fifthforce}

\appendix

\section{Screening Mechanisms}

The field equations of GR can be derived from the action
\begin{equation}
S_{\rm GR} = \frac{1}{16\pi G_{\rm N}}\int {\rm d}^4x\sqrt{-g}\ R + S_{\rm matter}[g_{\mu\nu}]\,, 
\label{eqn:GR_Action}
\end{equation}
where $g$ is the determinant of the metric tensor
$g_{\mu\nu}$, and $R$ is the Ricci scalar. The first term is the
Einstein-Hilbert action, while $S_{\rm matter}$ contains all matter
fields, with minimal couplings to $g_{\mu\nu}$. 

Modified gravity theories designed to explain cosmic acceleration
 generically reduce to scalar-tensor theories in certain limits. The addition of a 
 scalar field to the gravity theory leads to an additional attractive force. To produce
 interesting effects cosmologically while satisfying the tight local
 constraints on deviations from GR, these scalar fields must 
 be screened in the solar system. There are three known
 mechanisms for screening a scalar field -- the Vainshtein, chameleon
 and symmetron mechanisms -- all of which exploit the
 fact that the matter density at a typical location in the solar
 system is many orders of magnitude larger than the mean cosmic density.  
 The nonlinear equations obeyed by the scalar field are such that a large
 ambient density  results in its  
 decoupling from matter, thus shielding dense environments from the scalar. 
 In short, the three mechanisms rely on
 giving the scalar field 
a large mass (chameleon), a large inertia (Vainshtein), 
or by weakening its coupling to  matter (symmetron) 
\citep[see][for details on some of the simplified description here]{Hui09,Jain10}.
%(see Hui, Nicolis \& Stubbs 2009; 
%Jain \& Khoury 2010 for details on some of the simplified description that follows). 

By adding a suitable self-interaction potential $V(\varphi)$
to the action, the chameleon scalar field acquires mass which
is large in regions of high density,
thereby suppressing any long-range interactions. 
In addition to $V(\varphi)$ chameleon theories also have a 
coupling $A(\varphi)$ to matter fields - in fact a coupling arises when any scalar-tensor
theory is expressed in the Einstein frame:
\begin{equation}
S_{\rm cham} = \int {\rm d}^4x\sqrt{-g}\left(\frac{R}{16 \pi G} - \frac{1}{2}(\partial\varphi)^2 - V(\varphi)\right) + S_{\rm matter}\left[g_{\mu\nu}A^2(\varphi) \right] \,.
\label{eqn:Scham}
\end{equation}
In the Einstein frame, the scalar is not directly coupled to $R$, 
but due to the coupling of the metric to $A(\varphi)$ the motion of bodies in free fall 
is no longer along geodesics. 
%

%The equation of motion for $\varphi$ that derives from this action is
%
%$
%\Box\varphi = V_{,\varphi} - A^3(\varphi)A_{,\varphi} \tilde{T}\,,
%$
%
%where $\tilde{T} = \tilde{g}_{\mu\nu} \tilde{T}^{\mu\nu}$ is the trace of the energy-momentum tensor defined with respect to the
%Jordan-frame metric $\tilde{g}_{\mu\nu} = A^2(\varphi) g_{\mu\nu}$. Since matter fields couple %minimally to $\tilde{g}_{\mu\nu}$, this stress tensor is covariantly
%conserved: $\tilde{\nabla}_\mu  \tilde{T}^{\mu\nu} = 0$.

In the Newtonian regime, 
%We will also use the Taylor expansion 
%$A(\varphi)\simeq 1+\alpha\varphi$. 
using the energy density $\rho$ conserved in the Einstein frame, $\varphi$ obeys the equation: 
\begin{equation}
\nabla^2\varphi = V_{,\varphi} + A_{,\varphi}\rho 
%= V_{,\varphi} +\alpha \rho\,.
\label{eqn:phistat}
\end{equation}
Thus, because of its coupling to matter fields, the scalar field is affected by the 
ambient matter density.  Its profile is governed by an effective potential
\begin{equation}
V_{\rm eff}(\varphi) = V(\varphi) +A(\varphi)\rho \,.
\label{eqn:Veffcham}
\end{equation}
For suitably chosen $V(\varphi)$ and $A(\varphi)$, this effective potential
can develop a minimum  at some finite field value $\varphi_{\rm min}$ in
the presence of background matter density, where the mass of the
chameleon field $m_{\rm eff}^2 = V_{{\rm eff}{,\varphi\varphi}}(\varphi_{\rm min}) $
is sufficiently large to evade local constraints.

The condition for screening in chameleon theories is $\epsilon_{\rm chameleon}<1$ 
where the screening parameter is
\begin{equation}
\epsilon_{\rm chameleon} \simeq \left| \frac{\varphi_b/2\alpha}{\Phi_N} \right|
\end{equation}
where $\Phi_N$ is the Newtonian potential,
 $\varphi_b$ is the field value at cosmic mean density, 
and $\alpha$ is the coupling constant in $A(\varphi)$ 
(to be explicit, a general form of $A(\varphi)$ can be 
Taylor expanded as $A(\varphi)\simeq 1 + 2\alpha\varphi$). 
For all $f(R)$ models $\alpha=1/\sqrt{6}$, and 
the numerator $\varphi_b/2\alpha \approx f_{R0}$ where $f_{R0}$
is a parameter commonly used in the literature to constrain the model 
\citep{Hu07}. %(Hu \& Sawicki 2007). 
The strongest constraints from large-scale structure
come from cluster abundances and set an upper limit of $f_{R0} \approx 10^{-4}$
\citep{Schmidt09, Lombriser10}. %(Schmidt, Vikhlinin \& Hu 2009; Lombriser et al 2010). 
%For tests with dwarf 
%galaxies it is convenient to equate $\varphi_b/2\alpha$ to the value of the 
%Newtonian potential $\Phi_N^{\rm tran}$ at which halos transition from being
%screened to unscreened. 

We can write the force enhancement in chameleon theories in terms
of an additional potential due to the scalar field $\Phi_{\rm scalar}$ as
\begin{equation}
\Phi_{\rm total} = \Phi_N + \Phi_{\rm scalar} \approx
- \frac{G_N M}{r}\left[ 1 + \hat\alpha \ E(-r/\lambda_c) \right]
\label{eqn:Yukawa}
\end{equation} 
where the Compton wavelength $\lambda_c \equiv 1/m_{\rm eff}$, 
where $m_{\rm eff}$ is the effective mass of the scalar field. The function 
$E$  can be approximated as a constant $\delta$ times the exponential Yukawa suppression: 
$E(x) \equiv \delta e^x$. 
For distances much larger than $\lambda_c$, the fifth
force due to the scalar contribution is exponentially suppressed. 
The key feature of chameleon screening is that 
$\lambda_c$ is not constant but depends on the local mass distribution, 
in particular it is very small (less than a $mm$) inside objects with large
Newtonian potentials ($|\Phi_N| \gsim 10^{-6}$), causing the
scalar force to be strongly damped: such objects are screened. 
The exterior of such objects in a cosmological background  is described by 
$\delta\ll1$ since only a thin shell near the surface acquires the scalar charge. 
On the other hand objects that are unscreened have $\delta = 1$ and $\lambda_c$ large (equal
to the cosmological background value) both inside and outside. Thus dynamics within
and exterior to unscreened objects is well approximated through an enhanced 
gravitational constant:  $G_N \to G_N (1+\hat \alpha)$. Note that 
this simplified description does not capture the 
 behavior within the thin-shell, which may be relevant for some screened halos.

For the purposes of this study, 
we can describe the observational features of chameleon theories through two
parameters: 1. The coupling constant $\hat\alpha = \Delta G/G_N$ 
for unscreened objects; it is generally expected to be of order unity
 and is $\hat\alpha = 2\alpha^2 = 1/3$ for $f(R)$ models 
\citep[][note that it is generically expected to be larger 
 as the value 1/3 is not protected from quantum corrections]{Hui09}.
%(Hui, Nicolis \& Stubbs 2009 note that it is generically expected to be larger 
 %as the value 1/3 is not protected from quantum corrections). 
2. The value of the Newtonian
 potential $\Phi_N^{\rm tran}$ that provides the transition from screened to un-screened halos. 
 For $|\Phi_N | \ll | \Phi_N^{\rm tran}|$, $\delta \to 1$, so that forces are enhanced by 
 the factor $(1+\hat\alpha)$.  For screened objects on the other hand,  $|\Phi_N | \gg | \Phi_N^{\rm tran}|$,  $\delta\ll1$ and 
 Newtonian gravity is recovered.  An upper limit on $\Phi_N^{\rm
   tran}$ maps onto the theory parameter $\varphi_b$ for given
 $\alpha$. This is the parameter $f_{R0}$ in $f(R)$ theories. Thus 
constraints from unscreened galaxies and from cosmological scales are
straightforwardly related within a screening scenario. 

{\bf Symmetron screening:} The second mechanism for hiding a scalar is achieved with 
symmetron
fields, proposed recently by \citet{Hinterbichler10}. %Hinterbichler \& Khoury (2009). 
The symmetron Lagrangian is
qualitatively similar to that of chameleon models, but the mechanism
is different. The
screening in this case relies on a scalar field acquiring a vacuum
expectation value (VEV) that is small in high-density regions and
large in low-density regions. An essential ingredient is that the
coupling to matter is proportional to this VEV, so that the scalar
couples with gravitational strength in low-density environments, but
is decoupled and screened in regions of high density. This is achieved
through a symmetry-breaking potential, hence the name symmetron. 

The parameter that determines whether a solution is screened or not is
\begin{equation}
\epsilon_{\rm sym} 
\equiv \frac{M^2}{\rho R^2} = \frac{M_{\rm sym}^2/6 M_{\rm Pl}^2}{|\Phi_N|} \,.
\label{eqn:alp}
\end{equation}
where $M_{\rm sym}$ is a parameter of the theory. For objects
with $\epsilon_{\rm sym} \ll 1$ the resulting
symmetron-mediated force on a test particle 
is suppressed by $\epsilon_{\rm sym}$ compared to the gravitational force. 
On the other hand for objects with $\epsilon_{\rm sym} \gg 1$, 
the symmetron gives an ${\cal O}(1)$ correction to the 
gravitational force. This behavior is very similar to
chameleon screening. The quantitative difference is that 
symmetron screening for partially screened objects penetrates
deeper into the halo (Clampitt et al, in preparation). The bottom line for
astrophysical tests of gravity is that chameleon and symmetron signatures
occur in similar physical situations. 
 
{\bf Vainsthein Screening:} The third mechanism relies on the
scalar field having derivative interactions that become large in
regions of high density or in the vicinity of massive
objects \citep{Vainshtein72,Hinterbichler10}. 
Perturbations of the scalar in such regions acquire a large
kinetic term and therefore decouple from matter. Thus the scalar
screens itself and becomes invisible to experiments. This 
Vainshtein effect  is the screening mechanism that operates
successfully for brane-world modifications of
gravity such as the DGP model and galileon scalar field theories. 

For a spherical object, Vainshtein screening is effective inside the 
radius 
\begin{equation}
r_{\rm Vain} = (r_c^2r_{\rm Sch})^{1/3}\,,
\label{eqn:rvain}
\end{equation}
with $r_{\rm Sch}$ denoting the Schwarzschild radius of the source. At
short distances, $r\ll r_{\rm Vain}$, the extra force is suppressed 
compared to gravity by the factor 
$(r/r_{\rm Vain})^{3/2} \ll 1 $. The ratio of the $r_{\rm Vain}$ to 
the virial radius is larger than 1 and is constant with mass, since 
both radii scale as $M^{1/3}$. 
\begin{equation}
r_{\rm Vain} \gsim r_{\rm virial}, \ \ \  {\rm independent\  of\  mass}. 
\end{equation}
Thus the tests discussed may not be  not useful for Vainshtein theories as they
rely on the dynamics of disks well inside halos. We note however that the 
expectations of screening for spherical objects may be violated in the
presence of tidal fields, which can leave galaxies along filaments
of large-scale structure unscreened (R. Scoccimarro, private
communication). 
At large distances, $r \gg r_{\rm Vain}$,  the 
enhancement to Newtonian gravity is exactly as for an unscreened 
object in $f(R)$ theories: a factor of $1/3$. Indeed for galaxies or clusters
larger than the Milky Way, the infall of other galaxies at distances larger
than the virial radius will have stronger signatures from Vainshtein theories
since the extra force is not suppressed by the thin shell factor for massive hosts. 

\end{document}